\documentclass[prl,showpacs,twocolumn,
               amsmath,amssymb,superscriptaddress,
               floatfix,nofootinbib]{revtex4-1}

\usepackage[utf8]{inputenc}
\usepackage{amsmath,amssymb,amsbsy,amsfonts}
\usepackage{bm}
\usepackage{times}

\usepackage{graphicx}
\usepackage{float}
\usepackage{morefloats}
\usepackage{rotating}

\usepackage[caption=false]{subfig}
\usepackage{booktabs}
\usepackage{multirow}
\usepackage{tabularx}
\usepackage{adjustbox}
\usepackage{nicefrac}
\usepackage{slashed}
\usepackage{verbatim}
\usepackage[dvipsnames]{xcolor}
\usepackage{ragged2e}

\usepackage{hyperref}
\hypersetup{
	colorlinks	=true,
	urlcolor	=blue,
	linkcolor	=blue,
	citecolor	=blue,
	pdftitle	={},
	pdfauthor	={Jia-Ming Xie, Jun-Xu Lu, Li-Sheng Geng},
	pdfsubject	={Two-pole structures demystified: chiral dynamics at work}
	}


\begin{document}

\title{Off-shell Chiral Dynamics in the $\Lambda(1405)$ Resonance and $K^-p$ Femtoscopic Correlations}

\author{Jia-Ming Xie}
\affiliation{School of Physics, Beihang University, Beijing 102206, China}
\affiliation{Department of Physics, Graduate School of Science, The University of Tokyo, Tokyo 113-0033, Japan}

\author{Zhi-Wei Liu}
\email[Corresponding author: ]{liuzhw@buaa.edu.cn}
\affiliation{Institute for Advanced Study in Nuclear Energy \& Safety, College of Physics and Optoelectronic Engineering, Shenzhen University, Shenzhen 518060,
Guangdong, China}

\author{Jun-Xu Lu}
\email[Corresponding author: ]{ljxwohool@buaa.edu.cn}
\affiliation{School of Physics, Beihang University, Beijing 102206, China}

\author{Haozhao Liang}
\email[Corresponding author: ]{haozhao.liang@phys.s.u-tokyo.ac.jp}
\affiliation{Department of Physics, Graduate School of Science, The University of Tokyo, Tokyo 113-0033, Japan}
\affiliation{Quark Nuclear Science Institute, The University of Tokyo, Tokyo 113-0033, Japan}
\affiliation{RIKEN Center for Interdisciplinary Theoretical and Mathematical Sciences, Wako 351-0198, Japan}

\author{Li-Sheng Geng}
\email[Corresponding author: ]{lisheng.geng@buaa.edu.cn}
\affiliation{Sino-French Carbon Neutrality Research Center, \'{E}cole Centrale de P\'{e}kin/School
of General Engineering, Beihang University, Beijing 100191, China}
\affiliation{School of
Physics,  Beihang University, Beijing 102206, China}
\affiliation{Peng Huanwu Collaborative Center for Research and Education, Beihang University, Beijing 100191, China}
\affiliation{Southern Center for Nuclear-Science Theory (SCNT), Institute of Modern Physics, Chinese Academy of Sciences, Huizhou 516000, China}

\begin{abstract}
We present the first systematic investigation of the $S=-1$ meson--baryon interaction within a fully off-shell covariant unitarized chiral effective field theory framework up to next-to-leading order. In particular, we perform a detailed comparison with the widely used on-shell approximation. We find that the resulting scattering observables are very similar, thereby confirming the validity of key results obtained within the on-shell scheme. A notable advantage of the off-shell treatment, however, is the absence of unphysical left-hand cuts induced by the on-shell approximation. Employing the off-shell amplitudes, we compute the femtoscopic correlation functions for $K^-p$ and $\pi^\pm\Sigma^\mp$ pairs. The $K^-p$ correlation functions are found to be consistent with previously published results based on the on-shell approximation, with marginal differences attributed to slight variations in the descriptions of the scattering data. The $\pi^\pm\Sigma^\mp$ correlation functions are predicted for the first time, and are expected to provide valuable constraints on the nature of the $\Lambda(1405)$ resonance and the coupled-channel chiral dynamics of the $K^-p$ system.
\end{abstract}

\maketitle

\section{Introduction} 
\label{sec:intro}
Unveiling the nature of nonperturbative strong interactions remains one of the central challenges in theoretical particle and nuclear physics, primarily due to color confinement and spontaneous chiral symmetry breaking, characteristic of low-energy quantum chromodynamics (QCD)~\cite{Gross:1973id, Politzer:1973fx,Gross:1973ju, Gross:2022hyw}. In this context, antikaon-nucleon ($\bar{K}N$) scattering serves as a sensitive probe into the intricate meson-baryon dynamics in the strangeness $S=-1$ sector, offering insights with far-reaching implications for both hadronic and hypernuclear physics~\cite{Batty:1997zp,Hyodo:2011ur,Gal:2016boi,Curceanu:2019uph,Hyodo:2020czb}. Among the various phenomena in this domain, the $\Lambda(1405)$ resonance stands out due to its distinctive properties~\cite{Dalitz:1959dn,Kim:1965zzd} and its long-standing role as a benchmark~\cite{Callan:1985hy,Dalitz:1967fp,Veit:1984an,Veit:1984jr,Jennings:1986yg,Fink:1989uk} for testing theoretical frameworks beyond the conventional quark-model interpretation of baryon resonances~\cite{Capstick:1986ter,Klempt:2009pi}.

Over the past decades, unitarized chiral effective field theory~(U$\chi$EFT) has proven to be a powerful framework for investigating such systems, effectively combining the principles of chiral symmetry with coupled-channel unitarization techniques~\cite{Weinberg:1990rz,Kaiser:1995eg,Kaiser:1996js,Oset:1997it,Oller:2000ma,Hyodo:2011ur,Oller:2019opk}. This approach not only accounts for the observed features of the $\Lambda(1405)$ with notable consistency~\cite{Mai:2020ltx} but also has led to the revelation of its two-pole structure~\cite{Oller:2000fj,Jido:2003cb}. This insight has significantly reshaped our theoretical understanding of low-energy hadron-hadron interactions~\cite{Meissner:2020khl,Mai:2022eur,Geng:2026ywt}. In a recent work, it was shown that such a two-pole structure can be experimentally verified using charmonium decays into $\bar{\Lambda}\Sigma\pi$ and $\bar{\Lambda}(1520)\Sigma\pi$~\cite{He:2026mkf}.

However, two key issues remain unresolved. First and foremost, whether the qualitative picture of the two-pole structure of $\Lambda(1405)$---already verified in lattice QCD studies~\cite{BaryonScatteringBaSc:2023zvt,BaryonScatteringBaSc:2023ori} and consistently observed across leading-order~(LO)~\cite{Oller:2000fj}, next-to-leading order~(NLO)~\cite{Borasoy:2005ie,Borasoy:2006sr,Oller:2005ig,Oller:2006yh,Ikeda:2012au,Guo:2012vv,Mai:2012dt,Ramos:2016odk,Sadasivan:2022srs}, and next-to-next-to-leading order~(NNLO)~\cite{Lu:2022hwm} computations within the on-shell approximation---persists under more rigorous off-shell calculations~\cite{Revai:2017isg,Nieves:2001wt,Garcia-Recio:2002yxy,Mai:2012dt,Morimatsu:2019wvk}. This is closely tied to the validity of the on-shell approximation itself, which has been explicitly scrutinized only in the context of analytical tree-level interaction kernels~\cite{Oller:1997ti}, while a comprehensive off-shell treatment of the scattering equation~\cite{Mai:2012dt,Sadasivan:2018jig} is capable of avoiding the inherent deficiencies of the former approach---in particular its inability to handle left-hand cuts arising from complicated dynamics---thereby facilitating a more precise determination of the subthreshold amplitude profile relevant to resonance-pole extractions. Equally important is whether the underlying mechanism responsible for the two-pole structure---in particular, the chiral symmetry-breaking pattern~\cite{Xie:2023cej,Xie:2023jve,Xie:2025nnq} and SU(3) flavor symmetry breaking~\cite{Jido:2003cb,Guo:2023wes}---is fundamentally altered in an off-shell framework~\cite{Zhuang:2024udv,Pittler:2025upn}. 
Indeed, existing off-shell studies in the meson--meson~\cite{Nieves:1998hp,Nieves:1999bx, Altenbuchinger:2013gaa} and meson--baryon~\cite{Nieves:2000km,Bruns:2010sv} sectors have shown that solving the Bethe--Salpeter (BS) equation with fully off-shell chiral potentials can lead to improvements in various aspects of the description, yet introduces no qualitative change in the dynamically generated spectrum compared to the on-shell approximation.

In addition, explicitly off-shell scattering amplitudes are indispensable for applications extending beyond two-body systems, such as many-body~($A\ge3$) nuclear matter calculations~\cite{Dote:2008in,Dote:2008hw,Ikeda:2010tk,Bayar:2011qj,Hyodo:2011ur}. This necessity is most prominently reflected in the interpretation of $K^-p$ momentum correlation functions (CFs) from high-energy $pp$ collisions at the LHC~\cite{ALICE:2019gcn,ALICE:2022yyh,Kamiya:2019uiw,Encarnacion:2024jge}. As a novel tool complementing traditional scattering experiments~\cite{Morita:2016auo,Morita:2019rph,Kamiya:2019uiw,Liu:2022nec,Vidana:2023olz,Ikeno:2023ojl,Torres-Rincon:2023qll,Liu:2023uly,Liu:2023wfo,Molina:2023oeu,Albaladejo:2024lam,Sarti:2023wlg,Liu:2024uxn,Liu:2024nac,Feijoo:2024bvn,Encarnacion:2024jge,Liu:2025rci,Ge:2025put,Liu:2025oar,Ramos:2025ibe,Shen:2025qpj}, femtoscopic correlations probe the short-distance dynamics of the emitting source. This spatial sensitivity naturally raises the question of how the off-shell structure of the scattering amplitude, inherent in the short-range production process, may influence the resulting observables~\cite{Epelbaum:2025aan,Molina:2025lzw}. Consequently, evaluating a consistent off-shell treatment is essential to determine its potential role in refining the extraction of $\bar{K}N$ interaction properties from high-precision measurements.

In this work, we present the first fully off-shell covariant calculation of the $K^-p$ coupled-channel scattering process up to NLO within U$\chi$EFT. 
By deriving the perturbative Feynman amplitudes within chiral perturbation theory ($\chi$PT)~\cite{Weinberg:1978kz,Gasser:1983yg,Gasser:1984gg} as the interaction kernel and numerically solving the three-dimensionally reduced BS integral equation~\cite{Salpeter:1951sz} in the off-shell scheme, we obtain the non-perturbative scattering amplitudes required for the subsequent analysis.
The low-energy constants (LECs) and physical cutoffs are determined by a global fit to a comprehensive set of experimental data, including total cross sections for the final states $\{K^-p,\bar{K}^0n,\pi^-\Sigma^+,\pi^0\Sigma^0,\pi^+\Sigma^-,\pi^0\Lambda, \eta\Lambda\}$~\cite{Humphrey:1962zz, Kim:1965zzd, Sakitt:1965kh, Kittel:1966zz, Evans:1983hz, Ciborowski:1982et, CrystalBall:2001uhc}, threshold branch ratios $\{\gamma, R_c, R_n\}$~\cite{Nowak:1978au, Tovee:1971ga}, and the $K^-p$ scattering length~\cite{Meissner:2004jr, Bazzi:2011zj}. Utilizing the so-obtained interactions, we perform a comparative analysis of two-particle $K^-p$ and $\pi^\pm\Sigma^\mp$ CFs using both on-shell and off-shell formulations. We also examine how off-shell dynamics modify the CFs profiles, particularly at low relative momenta and near the $K^-p$ threshold, and to what extent these dynamical effects can be disentangled from production-related parameters such as the source radius $R$ and channel weights $w_j$.

The remainder of this paper is organized as follows. In Sec.~\ref{sec:formalism}, we present the theoretical framework, including the chiral interaction kernel up to NLO, the off-shell BS equation, and the formalism for constructing femtoscopic CFs. In Sec.~\ref{sec:results}, we discuss the fitting results, the extracted pole structures, and the comparative analysis of $K^-p$ and $\pi^\pm\Sigma^\mp$ correlation functions in both on-shell and off-shell schemes. Finally, in Sec.~\ref{sec:summary}, we summarize our main findings and provide an outlook for future studies.

\section{Formalism} 
\label{sec:formalism}

\subsection{Chiral Lagrangian up to $\mathcal{O}(p^2)$}
\label{subsec:formulae}

Based on $\chi$PT and two-body unitarization techniques~\cite{Oller:2019opk}, U$\chi$EFT provides a systematic framework for describing low-energy hadron interactions. In this approach, observables are expanded in powers of external momenta and light-quark masses relative to the QCD scale $\Lambda_{\mathrm{QCD}}$. The structure of the effective Lagrangian is strictly constrained by the chiral symmetry of QCD and its spontaneous and explicit breaking patterns.

The LO $\mathcal{O}(p)$ chiral Lagrangian describing the interactions of pseudo-Nambu-Goldstone bosons (pNGBs) with octet baryons is given by~\cite{Weinberg:1996kr}
\begin{align}
    \mathcal{L}_{1} &= \mathrm{Tr} \left[ \bar{\mathcal{B}} (i \slashed{\mathcal{D}} - M_0) \mathcal{B} \right] \nonumber \\
    &\quad + \frac{D}{2} \mathrm{Tr} \left( \bar{\mathcal{B}} \gamma^{\mu} \gamma_5 \{u_{\mu}, \mathcal{B}\} \right) + \frac{F}{2} \mathrm{Tr} \left( \bar{\mathcal{B}} \gamma^{\mu} \gamma_5 [u_{\mu}, \mathcal{B}] \right),
    \label{eq:LOP1}
\end{align}
where $M_0$ denotes the common mass of the octet baryons in the chiral limit, and $D, F$ are the axial coupling constants~\cite{Ratcliffe:1998su}. The fields of the ground-state octet baryons are collected in the $3\times 3$ flavor matrix $\mathcal{B}$ as
\begin{equation}
    \mathcal{B} = \begin{bmatrix}
    \frac{1}{\sqrt{2}}\Sigma^0 + \frac{1}{\sqrt{6}}\Lambda & \Sigma^+ & p \\
    \Sigma^- & -\frac{1}{\sqrt{2}}\Sigma^0 + \frac{1}{\sqrt{6}}\Lambda & n \\
    \Xi^- & \Xi^0 & -\frac{2}{\sqrt{6}}\Lambda
    \end{bmatrix}.
    \label{eq:Bfields}
\end{equation}

The Weinberg-Tomozawa (WT)~\cite{Weinberg:1966kf,Tomozawa:1966jm} interaction arises from the covariant derivative $\mathcal{D}_{\mu}$, defined as $\mathcal{D}_{\mu} \mathcal{B} = \partial_{\mu} \mathcal{B} + [\Gamma_{\mu}, \mathcal{B}]$, where the chiral connection $\Gamma_{\mu}$ and the axial-vector field $u_{\mu}$ are expressed in terms of the chiral fields $u$ as
\begin{align}
    \Gamma_{\mu} &= \frac{1}{2} \left( u^{\dagger} \partial_{\mu} u + u \partial_{\mu} u^{\dagger} \right), \\
    u_{\mu} &= i \left( u^{\dagger} \partial_{\mu} u - u \partial_{\mu} u^{\dagger} \right).
\end{align}
The field $u$ is the exponential representation of the pNGB matrix fields $\Phi$ as
    $u = \exp \left( i \Phi/\sqrt{2} f \right)$,
with the pNGB octet matrix $\Phi$ given by
\begin{equation}
    \Phi = \begin{bmatrix}
    \frac{1}{\sqrt{2}}\pi^0 + \frac{1}{\sqrt{6}}\eta_8 & \pi^+ & K^+ \\
    \pi^- & -\frac{1}{\sqrt{2}}\pi^0 + \frac{1}{\sqrt{6}}\eta_8 & K^0 \\
    K^- & \bar{K}^0 & -\frac{2}{\sqrt{6}}\eta_8
    \end{bmatrix}.
    \label{eq:Phifields}
\end{equation}
Here, the constant $f$ corresponds to the meson decay constant in the chiral limit at tree level~\cite{Pich:1995bw}.

At the NLO $\mathcal{O}(p^2)$, the chiral Lagrangian relevant to meson-baryon scattering is~\cite{Borasoy:2005ie,Borasoy:2006sr,Oller:2005ig,Oller:2006yh,Ikeda:2012au,Guo:2012vv,Mai:2012dt,Ramos:2016odk,Sadasivan:2022srs}
\begin{align}
    \mathcal{L}_{2} &= b_0 \mathrm{Tr}(\bar{\mathcal{B}}\mathcal{B}) \mathrm{Tr}(\chi_{+}) + b_D \mathrm{Tr}(\bar{\mathcal{B}}\{\chi_+, \mathcal{B}\}) \nonumber \\
    &\quad + b_F \mathrm{Tr}(\bar{\mathcal{B}}[\chi_+, \mathcal{B}]) + b_1 \mathrm{Tr}(\bar{\mathcal{B}}\{u_{\mu}, [u^{\mu}, \mathcal{B}]\}) \nonumber \\
    &\quad + b_2 \mathrm{Tr}(\bar{\mathcal{B}}[u_{\mu}, [u^{\mu}, \mathcal{B}]]) + b_3 \mathrm{Tr}(\bar{\mathcal{B}}u_{\mu}) \mathrm{Tr}(u^{\mu}\mathcal{B}) \nonumber \\
    &\quad + b_4 \mathrm{Tr}(\bar{\mathcal{B}}\mathcal{B}) \mathrm{Tr}(u^{\mu}u_{\mu}).
    \label{eq:LOP2}
\end{align}
Here, $b_0$, $b_D$, $b_F$, and $b_{i}$ ($i=1, \dots, 4$) are the LECs to be determined through a fitting procedure.
The scalar field $\chi_+ = u\chi^{\dagger}u + u^{\dagger}\chi u^{\dagger}$ introduces explicit symmetry breaking via the quark mass matrix $\chi = 2B_0 \mathbf{m} = 2B_0 \mathrm{diag}\{m_u, m_d, m_s\}$.

\subsection{Off-shell interaction kernel}
\label{subsec:off_V}

To illustrate the derivation of the off-shell potentials, we take the $S$-wave component of the WT term as a concrete example. The same conceptual framework and formal procedures are systematically applied to the Born terms and the NLO contributions, whose explicit $S$-wave projected expressions are collected in Appendix~\ref{app:amplitudes}, ensuring a consistent treatment of the off-shell dynamics across all interaction sectors.

Extracting the covariant derivative term $\mathrm{Tr}(\Bar{\mathcal{B}}(i\gamma^{\mu}[\Gamma_{\mu},\mathcal{B}]))$ from the LO chiral Lagrangian Eq.~(\ref{eq:LOP1}) and expanding the chiral connection $\Gamma_{\mu}$ to the lowest order in the meson field $\Phi$ as
    $\Gamma_{\mu}=(\Phi\partial_{\mu}\Phi-\partial_{\mu}\Phi\Phi)/4f^2$,
one obtains the following WT chiral Lagrangian as a four-point vertex~\cite{Oset:1997it}:
\begin{equation}
    \mathcal{L}^{\mathrm{WT}}=\frac{1}{4f^2}\mathrm{Tr}(\Bar{\mathcal{B}}i\gamma^{\mu}[\Phi\partial_{\mu}\Phi-(\partial_{\mu}\Phi)\Phi,\mathcal{B}]),
\end{equation}
from which the tree-level WT amplitude for the scattering process $\phi_i B^{\sigma}_i \rightarrow \phi_j B^{\sigma'}_j$ can be derived~\cite{Jido:2003cb} as
\begin{equation}
    \mathcal{V}^{\mathrm{WT}}_{ij}=-\frac{C_{ij}^{\mathrm{WT}}}{8f_i f_j}\Bar{u}(p_j)\gamma^{\mu}u(p_i)(p_{i\mu}+p_{j\mu}),
\end{equation}
where the trace has been evaluated in SU(3) flavor space to produce the coefficient matrix $C_{ij}^{\mathrm{WT}}$ in the particle basis~\cite{Borasoy:2005ie}. The four-momenta of the incoming and outgoing baryons (mesons) are denoted by $p_i$ and $p_j$ ($p_i$ and $p_j$), respectively.

Adopting the 4-component Dirac spinor with the normalization convention $N(\mathrm{p})=\sqrt{E(\mathrm{p})+M}$ and $E(\mathrm{p})=\sqrt{\mathrm{p}^2+M^2}$, the formula reads
\begin{align}
    u(p,s) &= N(\mathrm{p}) \begin{pmatrix} 1 \\ \frac{\boldsymbol{\sigma} \cdot \mathbf{p}}{E(\mathrm{p})+M} \end{pmatrix} \chi_s, \label{eq:u_spinor} \\
    \bar{u}(p,r) &= N(\mathrm{p}) \begin{pmatrix} \chi_r^{\dagger} & -\chi_r^{\dagger} \frac{\boldsymbol{\sigma} \cdot \mathbf{p}}{E(\mathrm{p})+M} \end{pmatrix}, \label{eq:ubar_spinor}
\end{align}
where $\chi_{s,r}$ are the two-component Pauli spinors. Returning to the perturbative off-shell WT amplitude $\mathcal{V}^{\mathrm{WT}}_{ij}$, which serves as the interaction kernel of the scattering equation, the transition potential is given by~\cite{Hyodo:2011ur}
\begin{align}
    \mathcal{V}^{\mathrm{WT}}_{ij} 
    &= -\frac{C_{ij}^{\mathrm{WT}}}{8f_if_j}N_j N_i (\chi_j^{\sigma'})^{T} \bigg[ \Omega_{ji}^- \nonumber \\
    &\quad + \Omega_{ji}^+ \frac{\mathbf{p}_j \cdot \mathbf{p}_i + i(\mathbf{p}_j \times \mathbf{p}_i) \cdot \boldsymbol{\sigma}}{(E_j+M_j)(E_i+M_i)} \bigg] \chi_i^{\sigma},
    \label{eq:WT_potential}
\end{align}
where the energy-dependent factors are defined as
\begin{equation}
    \Omega_{ji}^\pm = 2\sqrt{s} \pm (M_i + M_j) + (E_i - p_i^0) + (E_j - p_j^0).
\end{equation}
In deriving Eq.~\eqref{eq:WT_potential}, we have utilized the Pauli identity: $(\boldsymbol{\sigma} \cdot \mathbf{p}_j)(\boldsymbol{\sigma} \cdot \mathbf{p}_i) = \mathbf{p}_j \cdot \mathbf{p}_i + i(\mathbf{p}_j \times \mathbf{p}_i) \cdot \boldsymbol{\sigma}$. The off-shell amplitude simplifies to the on-shell case $\mathcal{V}^{\mathrm{WT-ON}}$ under the condition $p_{i,j}^0=E_{i,j}$.

Finally, we perform the partial-wave decomposition of $\mathcal{V}^{\mathrm{WT}}_{ij}$. For the $S$-wave part, the spin-flip term and the term $\mathbf{p}_j \cdot \mathbf{p}_i = \mathrm{p}_j \mathrm{p}_i \cos \theta$ vanish upon angular integration. Averaging over the baryon spins and applying the orthonormality $\chi^{\dagger}_j \chi_i = \delta_{\sigma' \sigma}$, the $S$-wave potential is obtained as
\begin{align}
    \mathcal{V}^{\mathrm{WT}}_{ij;0} & (\sqrt{s},\mathrm{p}_j,\mathrm{p}_i,p_j^0,p_i^0) \nonumber \\
    &= -\frac{C_{ij}^{\mathrm{WT}}}{8f_if_j}N_j N_i \Big[ 2\sqrt{s}-M_i-M_j \nonumber \\
    &\quad - (p_i^0+p_j^0-E_i-E_j) \Big].
    \label{eq:WT_S_wave}
\end{align}

\subsection{Unitarized scattering equation}
\label{subsec:BSeq}
In the strangeness $S=-1$ sector, the $S$-wave potential up to NLO is decomposed as
\begin{equation}
    \mathcal{V}_{ij}^{S}(\sqrt{s},\mathrm{p}_j,\mathrm{p}_i,p_j^0,p_i^0) = \mathcal{V}_{ij}^{\mathrm{WT}} + \mathcal{V}_{ij}^{\mathrm{Born}} + \mathcal{V}_{ij}^{\mathrm{NLO}},
\end{equation}
where $\mathcal{V}_{ij}^{\mathrm{WT}}$ denotes the WT term, $\mathcal{V}_{ij}^{\mathrm{Born}}$ represents the $s$- and $u$-channel Born contributions, and $\mathcal{V}_{ij}^{\mathrm{NLO}}$ accounts for the NLO contact interactions~\cite{Hyodo:2011ur}.

In energy regions governed by resonances such as the $\Lambda(1405)$, the perturbative chiral expansion breaks down~\cite{Weinberg:1990rz,Kaiser:1995eg,Oset:1997it}, necessitating the re-summation of an infinite subset of diagrams to restore  two-body elastic unitarity~\cite{Oller:1997ng,Oller:1998hw,Oller:1998zr}. This is achieved by solving the coupled-channel BS equation in momentum space. For the $S$-wave amplitude, while maintaining the consistency of the $0$-components of all momenta, the integral equation reads
\begin{align}
    \mathcal{T}_{ji}(\sqrt{s}&, p_j, p_i) = \mathcal{V}_{ji}(\sqrt{s}, p_j, p_i) + \sum_k \int^{\Lambda} \widetilde{d\mathrm{q}} \nonumber \\
    &\quad \times\mathcal{V}_{jk}(\sqrt{s}, p_j, p_{i}^0, \mathrm{q})\mathcal{G}_k(\mathrm{q}; \sqrt{s}) \mathcal{T}_{ki}(\sqrt{s}, \mathrm{q}, p_i, p_{j}^0), \label{eq:BSE_S_wave}
\end{align}
where $p_{j,i} = (p_{j,i}^0, \mathrm{p}_{j,i})$, $\widetilde{d\mathrm{q}}\equiv \mathrm{q}^2 d\mathrm{q}/2\pi^2$, and the intermediate propagator is defined as
\begin{equation}
    \mathcal{G}_k(\mathrm{q}; \sqrt{s}) = \frac{E_{k,1}+E_{k,2}}{2E_{k,1}E_{k,2}} \frac{1}{s-(E_{k,1}+E_{k,2})^2+i\epsilon}.
\end{equation}

To fully perform the 3-dimensional reduction of the above integral equation, we adopt a specific scheme~\cite{Lu:2021gsb} where the zero-components of the baryon four-momenta, $p_i^0$ and $p_j^0$, are fixed by the total c.m. energy $\sqrt{s}$,
\begin{equation}
    p_l^0 = \frac{s + M_l^2 - m_l^2}{2\sqrt{s}}, \quad (l = i, j).
\end{equation}
This framework ensures that the unitarized amplitude $T_{ji}$ is consistently treated off-shell, where the off-shell effects vanish in the limit $p_l^0 \to E_l$.

While non-perturbative unitarization inevitably introduces model-dependent parameters, their physical interpretation varies across different frameworks. In standard on-shell approximations, such dependence is typically encoded by subtraction constants in dimensionally regularized loop functions~\cite{Oller:2000fj,Borasoy:2005ie,Ikeda:2012au,Guo:2012vv}. In contrast, our approach explicitly retains the off-shell dynamics by employing a physical exponential form factor
\begin{equation}
    f_{\Lambda}(\mathrm{p}_j, \mathrm{p}_i) = \exp \left[ -(\mathrm{p}_j^4 + \mathrm{p}_i^4)/\Lambda^4 \right],
    \label{eq:form_factor}
\end{equation}
which regularizes the UV divergences while maintaining a transparent connection to the underlying short-range dynamics~\cite{Machleidt:2011zz,Epelbaum:2019zqc,Lu:2021gsb}.

\subsection{Femtoscopic correlation functions} 
Within the coupled-channel generalization of the Koonin-Pratt formalism~\cite{Koonin:1977fh,Pratt:1990zq,Albaladejo:2024lam}, the two-particle momentum CFs for a observable channel $i$ at relative momentum $\mathrm{p}$ are given by
\begin{equation}
C_i(\mathrm{p})=\sum_{j} w_j \int d^3r\, S_R^j(r)\,
\left|\Psi^{(-)}_{ji}(\mathrm{p},r)\right|^2,
\label{eq:coupledC}
\end{equation}
where $j$ labels the coupled intermediate channels that feed into the observable channel $i$. The factors $w_j$ denote the channel weights, and $S_R^j(r)$ is the source distribution for the relative separation $r$ of the emitted pair in channel $j$, characterized by a source size $R$.
The outgoing coupled-channel wave function $\Psi^{(-)}_{ji}(\mathrm{p},r)$ is constructed from the off-shell unitarized scattering $T$-matrix obtained above. In momentum representation, it can be expressed schematically as
\begin{equation}
\Psi^{(-)}_{ji}(\mathrm{p})=\delta_{ij}\,\Phi^{(-)}_{0i}(\mathrm{p})
+\int \widetilde{d\mathrm{q}}\, \mathcal{G}_j(\mathrm{q})\,\mathcal{T}_{ji}(\mathrm{q},\mathrm{p})\,\Phi^{(-)}_{0i}(\mathrm{q}),
\label{eq:psi_short}
\end{equation}
where $\Phi^{(-)}_{0i}$ denotes the outgoing asymptotic reference wave (including the appropriate long-range Coulomb distortion when applicable)~\cite{Encarnacion:2024jge}.

With the specified ingredients, the CFs could also have been calculated within the on-shell framework using the Koonin-Pratt (KP) formula~\cite{Koonin:1977fh,Pratt:1990zq,Vidana:2023olz,Ikeno:2023ojl,Torres-Rincon:2023qll,Feijoo:2024bvn,Khemchandani:2023xup,Li:2024tof,Albaladejo:2024lam,Liu:2024nac,Liu:2025oar}
\begin{align}
    C(\mathrm{p}) &= 1 + \int_{0}^{\infty} d^{3}r S_R^j(r) \nonumber \\
    &\quad \times \left[ \left| j_{0}(\mathrm{p}r) + \mathcal{T}(\sqrt{s}) \cdot \tilde{G}(r, \sqrt{s}) \right|^{2} - |j_{0}(\mathrm{p}r)|^{2} \right],
\end{align}
where $j_0(\mathrm{p}r)$ is the spherical Bessel function and $\mathrm{p} = \sqrt{s - (m + M)^2} \sqrt{s - (m - M)^2} / (2\sqrt{s})$ represents the c.m. momentum of the particle pair with the meson mass $m$ and baryon mass $M$. The quantity $\tilde{G}$ is defined as
\begin{equation}
\tilde{G}(r, \sqrt{s}) = \int_{0}^{\mathrm{q}_{\text{max}}} \widetilde{d\mathrm{q}}\mathcal{G}_k(\mathrm{q}; \sqrt{s})j_0(\mathrm{q}r).
\end{equation}
The scattering amplitude in the KP model is parametrized via the effective range expansion (ERE)~\cite{Bethe:1949yr}
\begin{equation}
f(\mathrm{p}) \approx \left( -\frac{1}{a_0} + \frac{1}{2}r_{\text{eff}} \mathrm{p}^2 - i\mathrm{p} \right)^{-1},
\end{equation}
where $a_0$ and $r_{\text{eff}}$ are the scattering length and effective range, respectively.

\begin{figure}[htpb]
    \centering
    \includegraphics[width=3.4in]{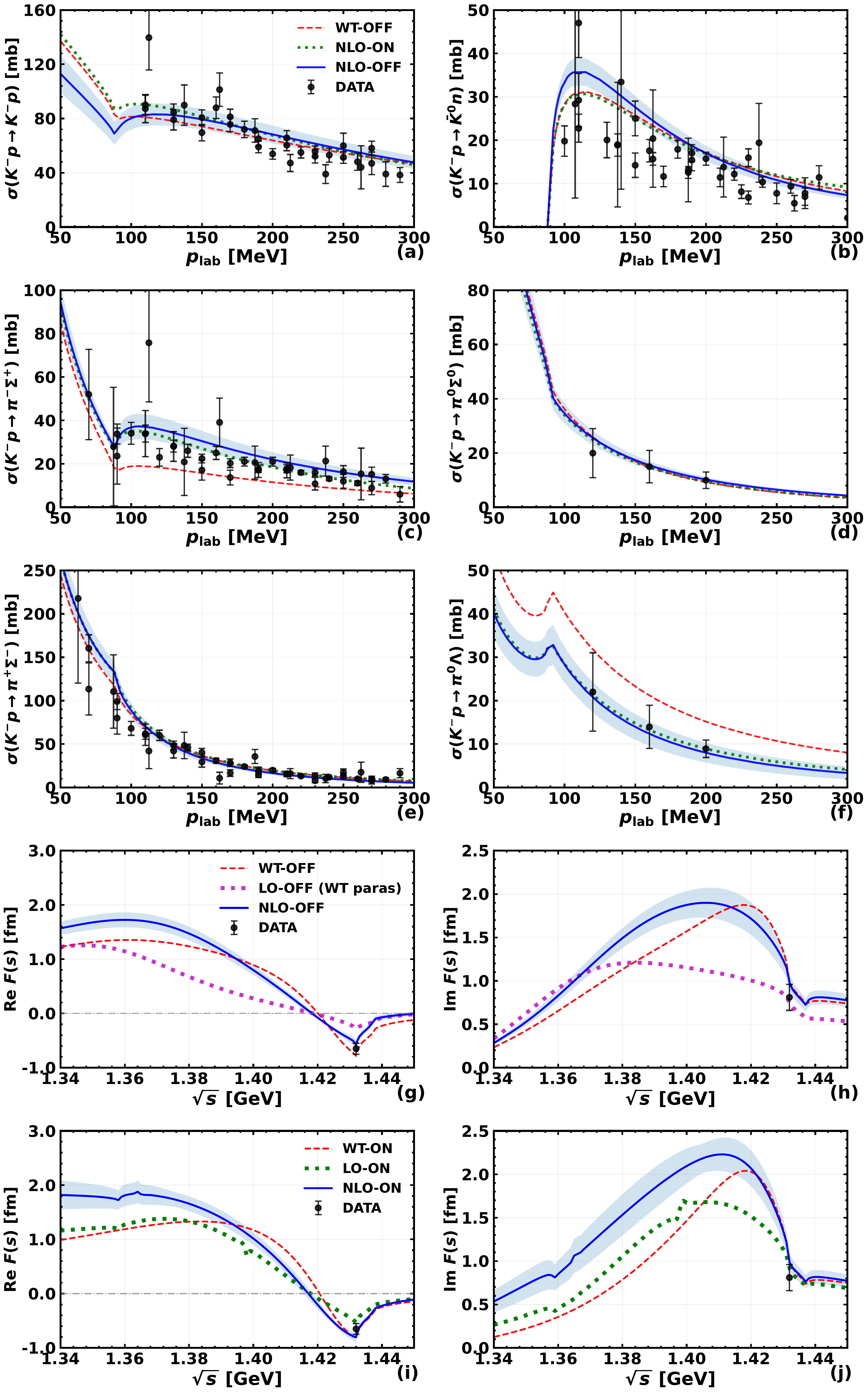}
   \caption{Global fit results for $K^-p$ scattering and subthreshold dynamics.
    The first six panels (a)--(f) present the comparison between experimental cross sections for various $S=-1$ channels $K^-p\rightarrow\{K^-p,\bar{K}^0n,\pi^-\Sigma^+,\pi^0\Sigma^0,\pi^+\Sigma^-,\pi^0\Lambda\}$~\cite{Humphrey:1962zz,Kim:1965zzd,Sakitt:1965kh,Kittel:1966zz,Evans:1983hz,Ciborowski:1982et} and the U$\chi$EFT results at both WT and NLO (including both off-shell and on-shell cases) with uncertainty bands estimated by the EFT Bayesian truncation~\cite{Furnstahl:2015rha,Melendez:2017phj,Melendez:2019izc}.
    The final four panels (g)--(j) illustrate the real and imaginary parts of the subthreshold $K^-p$ elastic amplitudes compared to data from Ref.~\cite{Bazzi:2011zj}. 
    Panels (g) and (h) show the off-shell results where the dotted lines represent our LO calculations performed using the WT parameter (paras) sets, while panels (i) and (j) provide the corresponding on-shell non-smooth amplitudes for comparison. 
    The smooth behavior of the full off-shell amplitudes, characterized by the absence of unphysical left-hand cut interferences, reveals that the two-pole structure of the $\Lambda(1405)$ is robust.}
    \label{fig:obs_off}
\end{figure}

\begin{table*}[htbp]
\centering
\caption{Off-shell determination of the threshold ratios $\{\gamma, R_c, R_n\}$ and scattering lengths $a_{K^-p}$ in comparison with available experimental data (EXP)~\cite{Nowak:1978au,Tovee:1971ga,Meissner:2004jr,Bazzi:2011zj}, $\chi^2/\text{d.o.f}$, and pole properties of the two $I=0$ states $\Lambda(1380)$ and $\Lambda(1405)$ obtained at WT and NLO.}
\label{tab:results_off}
\renewcommand{\arraystretch}{1.4}

\begin{tabular*}{\linewidth}{@{\extracolsep{\fill}} l c c c c c }
\hline\hline
& $\chi^2/\text{d.o.f}$ & $a_{K^-p}$ [fm] & $\gamma$ & $R_c$ & $R_n$ \\
\hline
NLO & 1.98 & $(-0.61 \pm 0.08) + i(0.99 \pm 0.10)$ & $2.36 \pm 0.24$ & $0.658 \pm 0.063$ & $0.184 \pm 0.023$ \\
WT & 4.11 & $-0.79 + i(0.99)$ & $2.38$ & $0.625$ & $0.231$ \\
EXP & -- & $(-0.65 \pm 0.10) + i(0.81 \pm 0.15)$ & $2.36 \pm 0.04$ & $0.664 \pm 0.011$ & $0.189 \pm 0.015$ \\
\hline
\end{tabular*}

\begin{tabular*}{\linewidth}{@{\extracolsep{\fill}} l c c c c c }
& Pole positions [MeV] & $|g_{\bar{K}N}|$ [GeV] & $|g_{\pi\Sigma}|$ [GeV] & $|g_{\eta\Lambda}|$ [GeV] & $|g_{K\Xi}|$ [GeV] \\
\hline
NLO $\Lambda(1380)$ & $(1375 \pm 3) - i(50 \pm 1)$ & $6.81 \pm 0.59$ & $7.52 \pm 0.66$ & $2.13 \pm 0.20$ & $0.61 \pm 0.14$ \\
NLO $\Lambda(1405)$ & $(1433 \pm 1) - i(28 \pm 2)$ & $8.15 \pm 0.45$ & $6.07 \pm 0.37$ & $3.52 \pm 0.19$ & $0.79 \pm 0.06$ \\
WT $\Lambda(1380)$ & $1365 - i(47)$ & $5.94$ & $6.59$ & $2.02$ & $1.36$ \\
WT $\Lambda(1405)$ & $1431 - i(21)$ & $3.62$ & $2.42$ & $1.90$ & $0.60$ \\
\hline\hline
\end{tabular*}

\end{table*}

\section{Results and Discussion}
\label{sec:results}
\subsection{Fitting strategy and parameter determination} 
Guided by the dominant role of the WT interaction at LO~\cite{Weinberg:1996kr}, we initiate our analysis by fitting the experimental data using the off-shell WT kernel. Our strategy is deliberately economical and symmetry-oriented~\cite{Oset:1997it}: a single cutoff parameter $\Lambda$ shared among all ten charge-neutral coupled channels is employed together with an average decay constant $f_{\mathrm{decay}}$ to effectively absorb the symmetry-breaking effects associated with the mass differences among the pseudoscalar meson octet~\cite{Guo:2012vv}. This choice contrasts with earlier studies that introduced six independent subtraction constants already at LO~\cite{Jido:2003cb}, as our motivation is to preserve SU(3) flavor symmetry as faithfully as possible~\cite{Oset:1997it, Oller:2000fj}. In a coupled-channel system with large threshold splittings, allowing channel-dependent cutoffs at LO risks obscuring the underlying dynamics through excessive fine-tuning~\cite{Xie:2023cej}.

To determine the LECs and cutoff parameters, we perform a global fit to the experimental data. 
Depending on the chiral expansion order, we employ two distinct statistical schemes to balance contributions from different datasets. 
For the optimization at the WT level (covering both off-shell and on-shell cases), we adopt a conventional definition of the $\chi^2$ per degree of freedom (d.o.f.)~\cite{Oset:1997it} as
\begin{equation}
\chi^{2}_{\text{d.o.f.}} = (N_{\text{tot}} - N_{\text{par}})^{-1} \sum_{j=1}^{J} \chi_{j}^{2},
\end{equation}
where $J$ is the total number of independent datasets, $N_{\text{tot}} = \sum_{j=1}^{J} M_{j}$ is the total number of experimental data points with $M_j$ being the number of data points in the $j$-th dataset, and $N_{\text{par}}$ denotes the number of free parameters. The individual $\chi_j^2$ for each dataset is defined as 
\begin{equation}
\chi_{j}^{2} = \sum_{i=1}^{M_{j}} (\mathcal{Y}_{j,i}^{\text{calc}} - \mathcal{Y}_{j,i}^{\text{data}} )^{2} / \sigma_{j,i}^{2},
\end{equation}
where $\mathcal{Y}_{j,i}^{\text{calc}}$ and $\mathcal{Y}_{j,i}^{\text{data}}$ are the theoretical and experimental values of the $i$-th data point in the $j$-th dataset, respectively, and $\sigma_{j,i}$ is the corresponding experimental uncertainty.
Since a minimal two-parameter $\{\Lambda, f_{\mathrm{decay}}\}$ description at the WT level cannot simultaneously reproduce all high-precision threshold ratios, we adopt these as the conventional error-weighting scheme that prioritizes the global energy dependence of scattering cross sections~\cite{Humphrey:1962zz, Sakitt:1965kh}.
This ensures that the fitted parameters capture the essential dynamics of the $\bar{K}N$ interaction rather than experimental fluctuations.

In the NLO analysis, where the data points are unevenly distributed among the $J$ measurements, we implement a cluster-weighting procedure to prevent data-intensive channels from statistically overwhelming crucial sparse observables~\cite{Garcia-Recio:2002yxy, Borasoy:2005ie, Ikeda:2011pi} 
\begin{equation}
\chi^{2}_{\text{d.o.f.}} = N_{\text{tot}} [J (N_{\text{tot}} - N_{\text{par}})]^{-1} \sum_{j=1}^{J} \chi_{j}^{2}/M_{j}.
\end{equation}
In this framework, the constraint of a universal cutoff is relaxed to six channel-dependent cutoffs, complemented by seven independent LECs $\{b_0, b_D, b_F, b_{1,2,3,4}\}$.
To enhance sensitivity to low-energy dynamics, we further adopt a refined error-weighting scheme~\cite{Borasoy:2006sr} that assigns larger weights to high-precision threshold observables, including the SIDDHARTA kaonic hydrogen data~\cite{Bazzi:2011zj} and threshold ratios~\cite{Nowak:1978au}.
These constraints strongly restrict the amplitudes near the $\bar{K}N$ threshold, leading to an improved description of the subthreshold region and the pole structure of the $\Lambda(1405)$.

The resulting global fit describes the meson--baryon scattering data well in most channels shown in Fig.~\ref{fig:obs_off}, with a mild discrepancy remaining in the $K^-p \to \eta\Lambda$ channel at higher energies~(see Appendix~\ref{app:paras}), which is known to be better described at NNLO~\cite{Lu:2022hwm}. 
To determine the LECs and cutoff parameters, we perform a global fit to the experimental data. 
Depending on the chiral expansion order, we employ two distinct statistical schemes to balance contributions from different datasets. 
For the optimization at the WT level (covering both off-shell and on-shell cases), we adopt a conventional definition of the $\chi^2$ per degree of freedom (d.o.f.)~\cite{Oset:1997it} as
\begin{equation}
\chi^{2}_{\text{d.o.f.}} = (N_{\text{tot}} - N_{\text{par}})^{-1} \sum_{j=1}^{J} \chi_{j}^{2},
\end{equation}
where $J$ is the total number of independent datasets, $N_{\text{tot}} = \sum_{j=1}^{J} M_{j}$ is the total number of experimental data points with $M_j$ being the number of data points in the $j$-th dataset, and $N_{\text{par}}$ denotes the number of free parameters. The individual $\chi_j^2$ for each dataset is defined as 
\begin{equation}
\chi_{j}^{2} = \sum_{i=1}^{M_{j}} (\mathcal{Y}_{j,i}^{\text{calc}} - \mathcal{Y}_{j,i}^{\text{data}} )^{2} / \sigma_{j,i}^{2},
\end{equation}
where $\mathcal{Y}_{j,i}^{\text{calc}}$ and $\mathcal{Y}_{j,i}^{\text{data}}$ are the theoretical and experimental values of the $i$-th data point in the $j$-th dataset, respectively, and $\sigma_{j,i}$ is the corresponding experimental uncertainty.
Since a minimal two-parameter $\{\Lambda, f_{\mathrm{decay}}\}$ description at the WT level cannot simultaneously reproduce all high-precision threshold ratios, we adopt these as the conventional error-weighting scheme that prioritizes the global energy dependence of scattering cross sections~\cite{Humphrey:1962zz, Sakitt:1965kh}.
This ensures that the fitted parameters capture the essential dynamics of the $\bar{K}N$ interaction rather than experimental fluctuations.

In the NLO analysis, where the data points are unevenly distributed among the $J$ measurements, we implement a cluster-weighting procedure to prevent data-intensive channels from statistically overwhelming crucial sparse observables~\cite{Garcia-Recio:2002yxy, Borasoy:2005ie, Ikeda:2011pi} 
\begin{equation}
\chi^{2}_{\text{d.o.f.}} = N_{\text{tot}} [J (N_{\text{tot}} - N_{\text{par}})]^{-1} \sum_{j=1}^{J} \chi_{j}^{2}/M_{j}.
\end{equation}
In this framework, the constraint of a universal cutoff is relaxed to six channel-dependent cutoffs, complemented by seven independent LECs $\{b_0, b_D, b_F, b_{1,2,3,4}\}$.
To enhance sensitivity to low-energy dynamics, we further adopt a refined error-weighting scheme~\cite{Borasoy:2006sr} that assigns larger weights to high-precision threshold observables, including the SIDDHARTA kaonic hydrogen data~\cite{Bazzi:2011zj} and threshold ratios~\cite{Nowak:1978au}.
These constraints strongly restrict the amplitudes near the $\bar{K}N$ threshold, leading to an improved description of the subthreshold region and the pole structure of the $\Lambda(1405)$.

\subsection{Subthreshold amplitudes and the two-pole structure} 
As shown in the final four panels of Fig.~\ref{fig:obs_off}, the real and imaginary parts of the subthreshold amplitudes exhibit distinct energy-dependent profiles. 
A notable advantage of the present off-shell scheme is its smooth analytic behavior in the subthreshold region. Unlike the on-shell approximation, which frequently suffers from unphysical distortions induced by left-hand cut interferences~\cite{Borasoy:2005ie,Pittler:2025upn}, our framework remains free from such artifacts.
This feature is essential for a reliable determination of pole positions and demonstrates a controlled convergence from the WT interaction to the full NLO kernel~\cite{Borasoy:2005ie,Mai:2022eur,Pittler:2025upn}.

\begin{figure}[h]
  \centering
  \includegraphics[width=\columnwidth]{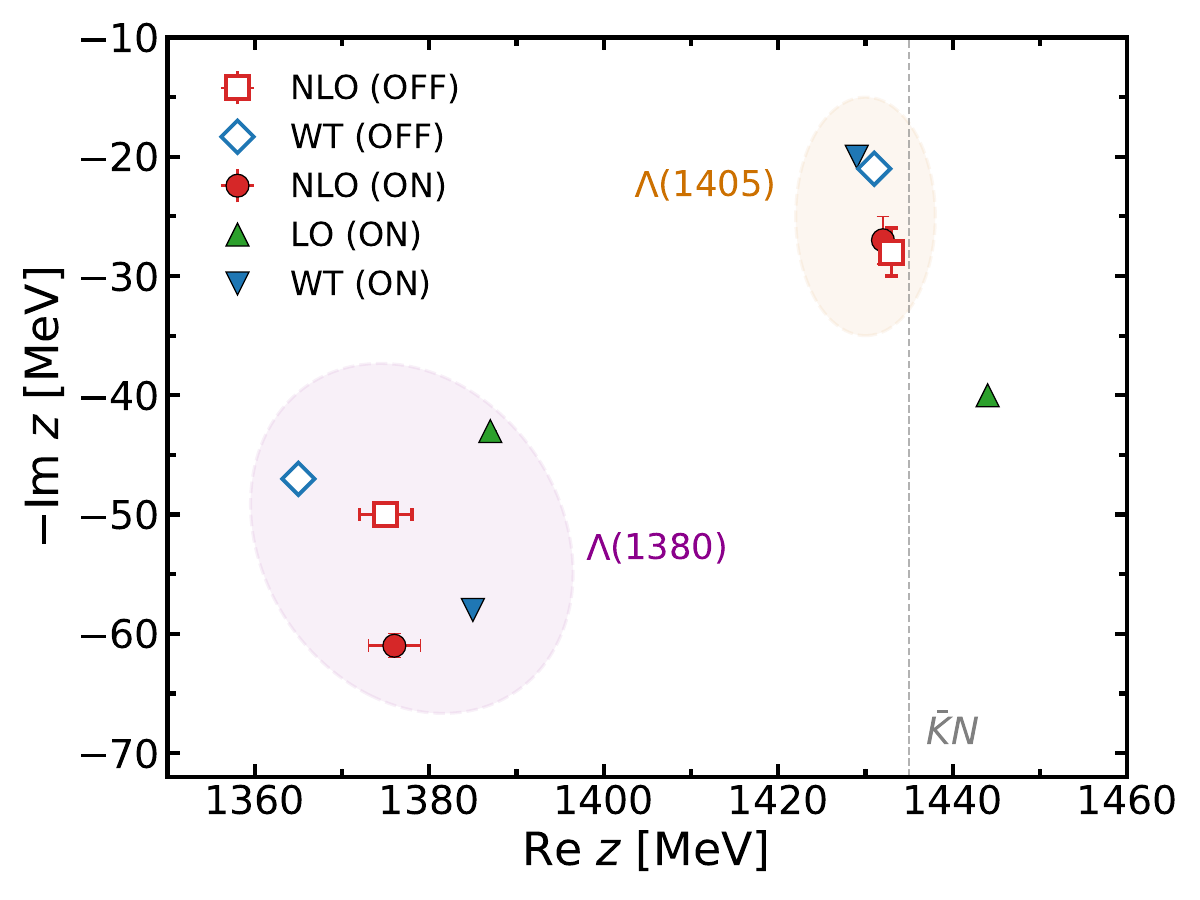}
  \caption{Pole positions of the $\Lambda(1380)$ and $\Lambda(1405)$ in the complex energy plane, obtained from NLO, LO, and WT schemes with both off-shell (OFF) and on-shell (ON) schemes. The dashed ellipses serve as visual guides to indicate the two-pole cluster structure.}
  \label{fig:pole_positions}
\end{figure}

Within the off-shell framework, the role of the Born terms differs qualitatively from that in conventional on-shell approaches, where they are often assumed to be suppressed~\cite{Oset:1997it,Jido:2003cb,Mai:2012dt}. An off-shell LO fit, including both WT and Born terms, yields a poorer description than the WT-only case, in contrast to the on-shell formulation where the Born terms bring marginal improvement (see Appendix~\ref{app:fit_on}). Moreover, applying the WT-optimized parameters to the full LO kernel reveals substantial deviations in the subthreshold amplitudes, indicating that the off-shell momentum dependence of the Born terms induces nontrivial dynamics.
This observation implies that, in an off-shell formulation, the Born terms should be treated consistently together with the NLO contact interactions. Grouping these contributions at NLO leads to a more stable and physically meaningful description of the coupled-channel dynamics without compromising the LO SU(3) symmetry.

Employing a unified framework that consistently incorporates off-shell dynamics, we extract the two-pole structure of the $\Lambda(1405)$ shown in Table~\ref{tab:results_off} at both WT and NLO from the second Riemann sheet.
The couplings of the poles to various channels $i,j$ are obtained from the residues of the poles on the complex plane as $T_{ij}(s)=\lim_{s\rightarrow s_R}g_ig_j/(s-s_R)$. The resulting pole positions are stable against changes in chiral order, demonstrating that the two-pole nature is an intrinsic chiral dynamical feature rather than an artifact of on-shell extrapolations~\cite{Meissner:2020khl,Lu:2022hwm,Xie:2023cej}.
For comparison, we also present all pole positions in the complex energy plane in Fig.~\ref{fig:pole_positions}.
The lower pole $\Lambda(1380)$ is broadly consistent across all schemes, while the higher pole $\Lambda(1405)$ exhibits a notable sensitivity to the treatment of on-shell approximation, particularly at LO, where it appears as a virtual state above the $\bar{K}N$ threshold.

In effective field theory calculations, theoretical uncertainties arise from the truncation of the chiral expansion at a given order. We estimate these uncertainties following the Bayesian approach developed in Refs.~\cite{Furnstahl:2015rha,Melendez:2017phj,Melendez:2019izc,Epelbaum:2019zqc} and generally utilized in Refs.~\cite{Lu:2021gsb,Lu:2022hwm}. 
The expansion parameter is defined as
    $Q = m_{\mathrm{ave}}/\Lambda_b$,
where $m_{\mathrm{ave}} \approx 370$~MeV represents the average meson mass in the $\bar{K}N$ system, and $\Lambda_b \approx 1.16$~GeV is the breakdown scale of the chiral expansion, yielding $Q \approx 0.32$.
For an observable $X$ computed at the NLO, the truncation error is estimated as
    $\Delta X_{\mathrm{NLO}} = |X_{\mathrm{LO}}| \times Q^2 \times \max\left(1, \left|\tilde{c}_1\right|\right)$,
where the coefficient $\tilde{c}_1$ characterizes the relative size of the NLO correction determined via the Bayesian approach.
The factor $\max(1, |\tilde{c}_1|)$ ensures that the error estimate remains conservative when the observed NLO correction exceeds the naive $\mathcal{O}(Q)$ expectation.

\begin{figure}[htbp]
    \centering

    \begin{minipage}[b]{0.5\linewidth}
        \includegraphics[width=\linewidth]{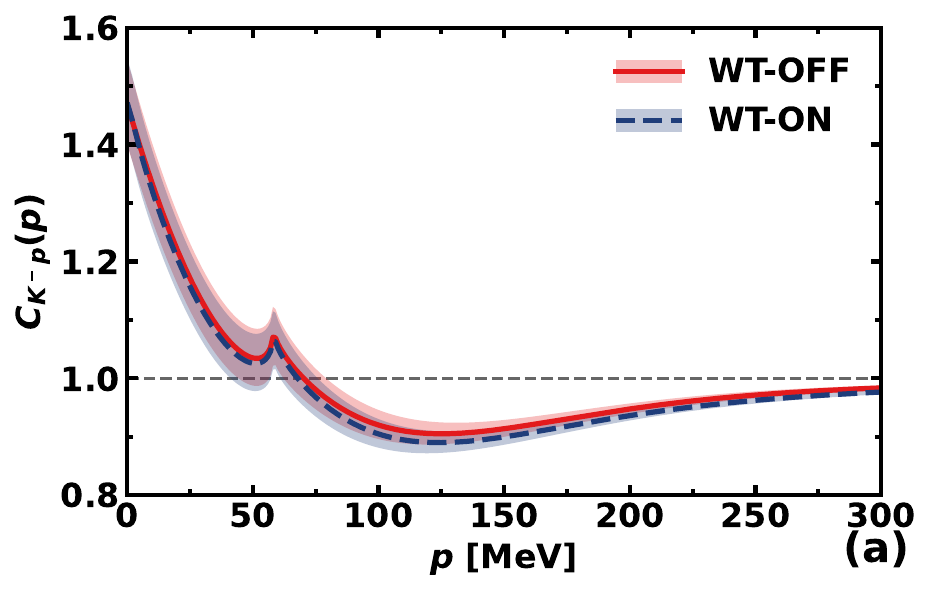}
    \end{minipage}
    \hspace{-0.74em}
    \begin{minipage}[b]{0.5\linewidth}
        \includegraphics[width=\linewidth]{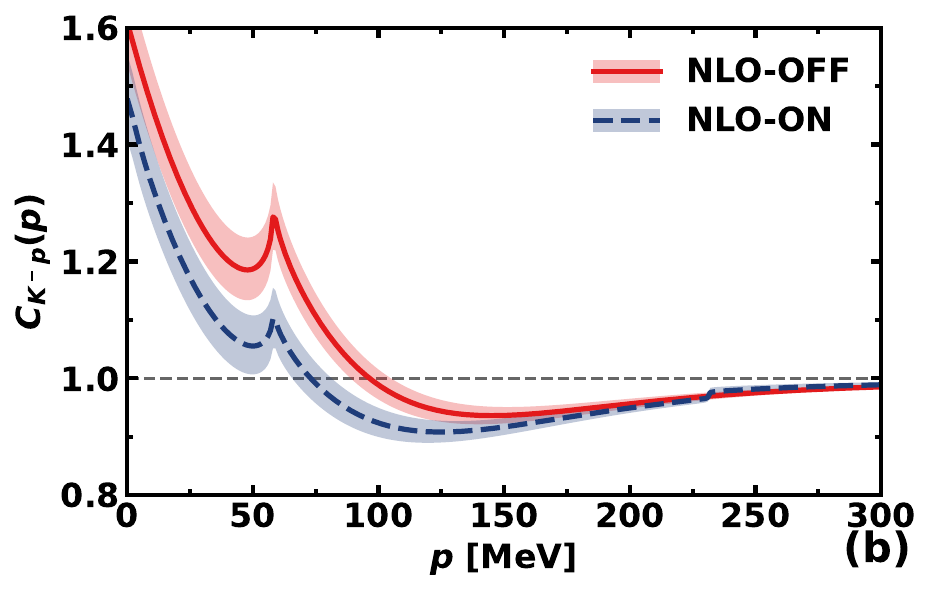}
    \end{minipage}


    \begin{minipage}[b]{0.5\linewidth}
        \includegraphics[width=\linewidth]{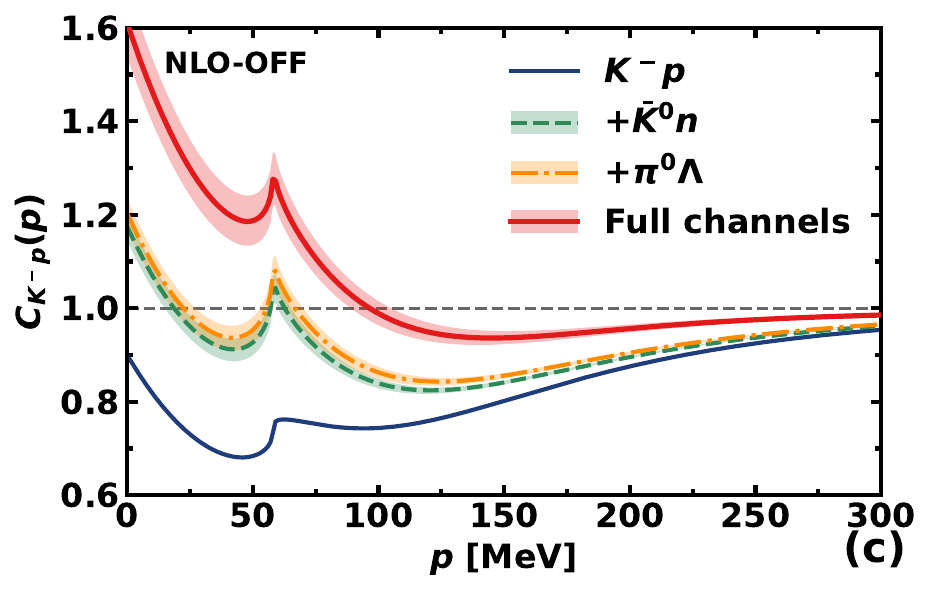}
    \end{minipage}
    \hspace{-0.74em}
    \begin{minipage}[b]{0.5\linewidth}
        \includegraphics[width=\linewidth]{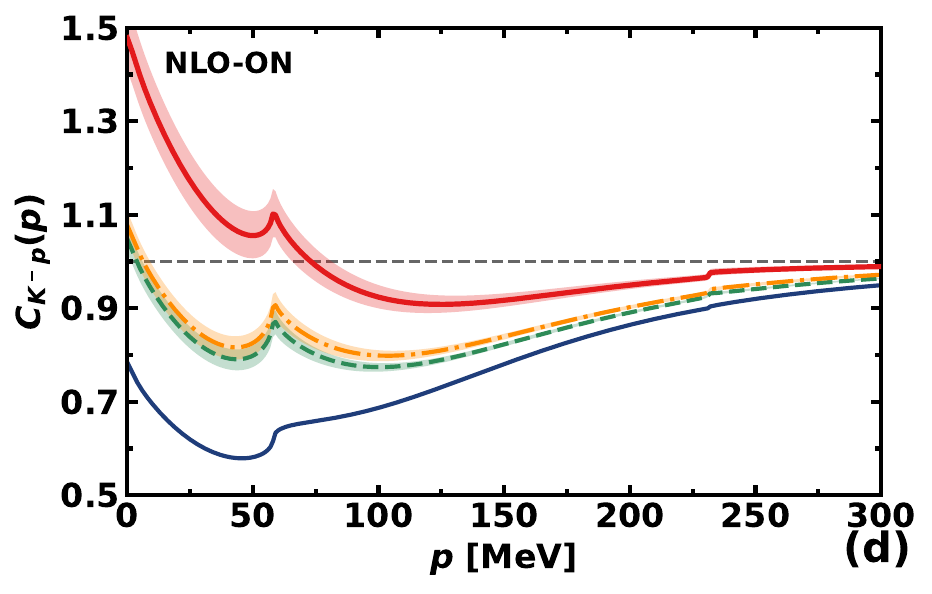}
    \end{minipage}
    \caption{$K^-p$ CFs and the impact of off-shell dynamics for $R = 1.0$~fm.
    (a)--(b): Off-shell versus on-shell comparison at fixed chiral order, showing effective reabsorption of off-shell effects at WT but noticeable deviations at NLO. 
    (c)--(d): Channel-by-channel contributions to the total CFs for the NLO off-shell and on-shell cases, demonstrating that the $K^-p$ single-channel contributions already exhibit a marked discrepancy even in the vicinity of the threshold.
    The shaded error bands for the CFs in panels (a)–(d) reflect the uncertainties originating from the channel weights $w_j$~\cite{Encarnacion:2024jge}.}
    \label{fig:CFR1p0}
\end{figure}

\begin{table}[h]
\centering
\caption{Scattering length $a_{K^-p}$ and the corresponding reduced chi-squared ($\chi^2/\mathrm{d.o.f.}$) for the $S=-1$ sector: comparison between off-shell and on-shell schemes.}
\label{tab:akp_fits} 
\setlength{\tabcolsep}{8pt}
\renewcommand{\arraystretch}{1.2}
\begin{tabular}{lcc} 
\hline\hline
Model & $\chi^2/\mathrm{d.o.f.}$ & $a_{K^-p}$ [fm] \\
\hline
NLO-OFF & 1.98 & $(-0.61 \pm 0.08) + i(0.99 \pm 0.10)$ \\
WT-OFF  & 4.11 & $-0.79 + i(0.99)$ \\
EXP & --   & $(-0.65 \pm 0.10) + i(0.81 \pm 0.15)$ \\
\hline
NLO-ON  & 1.82 & $(-0.81 \pm 0.08) + i(1.00 \pm 0.10)$ \\
LO-ON   & 3.18 & $-0.55 + i(0.93)$ \\
WT-ON   & 3.48 & $-0.81 + i(1.00)$ \\
\hline\hline
\end{tabular}
\end{table}

\subsection{Femtoscopic correlation functions}

Using the fitted interactions, we compute the $K^-p$ CFs assuming a Gaussian source with a typical radius $R=1.0~\mathrm{fm}$. The channel weights $w_j$ are taken from the Valencia (VLC) method proposed in Ref.~\cite{Encarnacion:2024jge}.
At the WT level, the off-shell and on-shell CFs are nearly identical [Fig.~\ref{fig:CFR1p0}(a)], indicating that off-shell effects can be effectively reabsorbed into the fitted parameters $\{\Lambda, f_{\mathrm{decay}}\}$. This naive renormalization, however, is slightly modified at the NLO level. Some visible variations appear due to the momentum dependence of the Born and NLO contact terms [Fig.~\ref{fig:CFR1p0}(b)].

However, the noticeable discrepancy between the NLO off-shell and on-shell results in the low-$p$ near-threshold region [Fig.~\ref{fig:CFR1p0}(b)] cannot be attributed to a direct manifestation of explicit off-shell effects~\cite{Epelbaum:2025aan,Molina:2025lzw}. 
While both schemes achieve a comparable quality of fit to the experimental data in Table~\ref{tab:akp_fits}, their overall consistency is difficult to equate.
Furthermore, whether the coupled-channel effects behave consistently across these two dynamical models remains an open question.

\begin{figure}[htbp]
    \centering
    \includegraphics[width=0.48\textwidth]{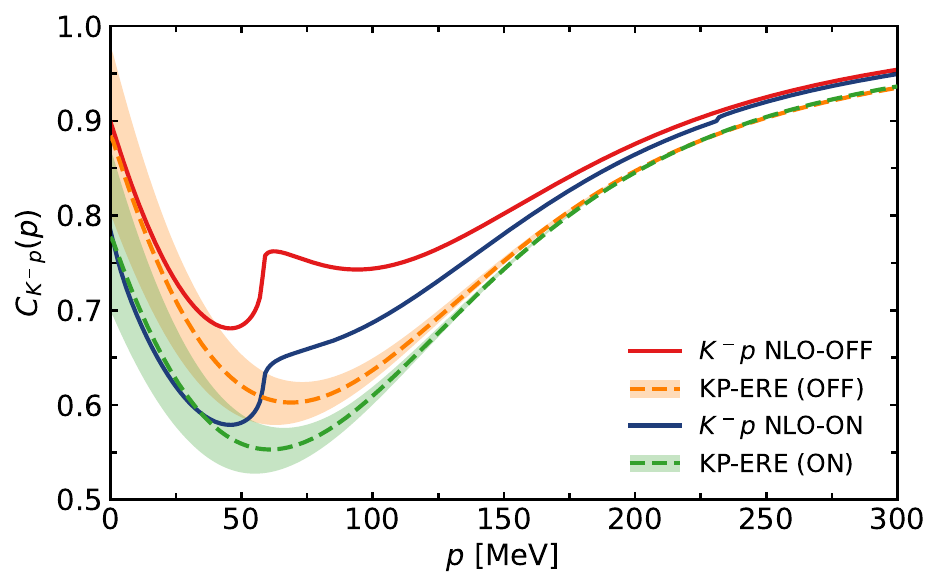}
    \caption{Comparison of the $K^-p$ single-channel CF contributions between NLO off-shell and on-shell schemes, alongside the on-shell KP formula results with ERE; the results show that the near-threshold discrepancy between the off-shell and on-shell CFs arises solely from the different scattering lengths $a_{K^-p}$ rather than explicit off-shell effects. The shaded error bands for the CFs arise from the uncertainties of the $a_{K^-p}$ in Table~\ref{tab:akp_fits}.}
    \label{fig:CF_PWA_vs_CKP}
\end{figure}

To address this, we first decompose the contributions of individual channels to the $K^-p$ CFs in Figs.~\ref{fig:CFR1p0}(c) and (d). 
It is found that even when considering only the $K^-p$ single-channel wave-function term, a substantial difference persists between the off-shell and on-shell scenarios near threshold.
Subsequently, we examine the specific contribution of explicit off-shell effects within the $K^-p$ single-channel CFs. 
By utilizing the scattering lengths $a_{K^-p}$ in Table~\ref{tab:akp_fits}, we approximate the on-shell scattering amplitudes via the effective range expansion (ERE). 
Incorporating these into the on-shell single-channel KP formula~\cite{Koonin:1977fh,Pratt:1990zq,Vidana:2023olz,Ikeno:2023ojl,Torres-Rincon:2023qll,Feijoo:2024bvn,Khemchandani:2023xup,Li:2024tof,Albaladejo:2024lam,Liu:2024nac,Liu:2025oar} allows us to accurately mimic the results of the rigorous calculations at low momentum. 
Fig.~\ref{fig:CF_PWA_vs_CKP} demonstrates that the discrepancy between the NLO off-shell and on-shell results near the threshold is determined by the different fits to the experimental data, i.e., the different scattering length $a_{K^-p}$ predicted.

\begin{figure}[htbp]
    \centering
    \includegraphics[width=0.48\textwidth]{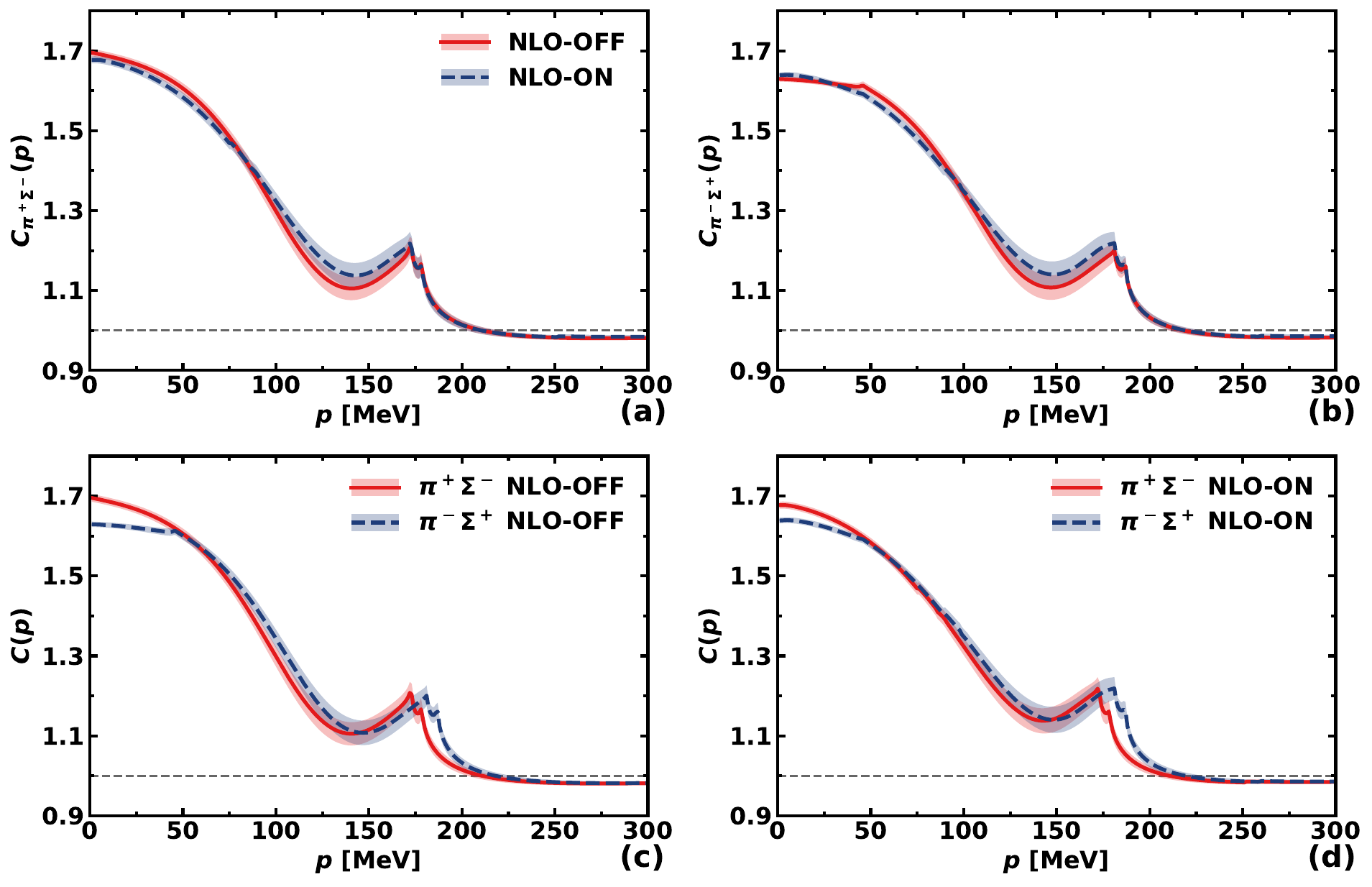}
    \caption{The $\pi^{\pm}\Sigma^{\mp}$ CFs with NLO-OFF and NLO-ON for $C_{\pi^+\Sigma^-}(p)$~(a) and $C_{\pi^-\Sigma^+}(p)$~(b). Panels~(c) and~(d) regroup the same curves by scheme to highlight the fine structure between the two channels.}
    \label{fig:CF2_NLO}
\end{figure}

The discussion above regarding the differences in the CFs between the NLO off-shell and on-shell cases is further corroborated by Fig.~\ref{fig:CF2_NLO}. As can be seen from panels (a) and (b), the CFs of $\pi^+\Sigma^-$ and $\pi^-\Sigma^+$ do not exhibit a sufficient observable difference between the off-shell and on-shell cases; the two are even indistinguishable, unlike what is shown in the $K^-p$ CFs. This once again demonstrates that, with our fitted parameters, the off-shell effects do not manifest any observable dynamical signatures. 
It is worth noting that our off-shell treatment of the non-perturbative scattering amplitude still demonstrates a necessary improvement over the on-shell approximation in the CFs, which is explicitly manifested in the $\pi^+\Sigma^-$ and $\pi^-\Sigma^+$. As illustrated in Fig.~\ref{fig:CF2_NLO}, the former on-shell case exhibits a discontinuity in the CFs at $p_{\rm cm} \approx 75~\mathrm{MeV}$, while the latter appears at $p_{\rm cm} \approx 90~\mathrm{MeV}$. This behavior is analogous to the discontinuity observed in panels (i) and (j) of Fig.~\ref{fig:obs_off} for the $K^-p$ subthreshold amplitude at $\sqrt{s} \approx 1.36~\mathrm{GeV}$, and originates from the left-hand cut introduced by the $u$-channel $\Lambda$ exchange in the $\eta\Lambda \to \eta\Lambda$ chiral perturbative amplitude~\cite{Borasoy:2005ie,Pittler:2025upn}. Nevertheless, such discontinuities do not lead to any substantial difference in the resulting CFs. To better visualize the fine structure, we regroup the results by scheme in Figs.~\ref{fig:CF2_NLO}(c) and~\ref{fig:CF2_NLO}(d), where the $\pi^+\Sigma^-$ and $\pi^-\Sigma^+$ CFs are overlaid within the NLO-OFF and NLO-ON scenarios, respectively. A visible splitting between the two charge channels emerges near the $K^-p$ and $\bar{K}^0 n$ threshold cusps, reflecting the isospin-breaking effect in the $\pi\Sigma$ final-state interaction.

\subsection{Comparison with the $pp$ collision data}

To investigate the experimental signatures of off-shell dynamics, we fit the ALICE $K^-p$ CFs~\cite{ALICE:2019gcn} following the strategy of Refs.~\cite{Kamiya:2019uiw,ALICE:2022yyh}, with the key difference that the source size $R$ is treated here as a free parameter~\cite{Molina:2025lzw}.
The experimental CF is analyzed using the following fitting formula involving the pair purity parameter $\lambda$ and a normalization constant $\mathcal{N}$ as
$C_{\mathrm{fit}}(k^*) = \mathcal{N}[1+\lambda\{C(k^*)-1\}]$,
where $C(k^*)$ denotes the theoretically calculated CF. The fitting range is restricted to low momenta $p \lesssim 120$ MeV/c to avoid the kinematic region dominated by the resonance contribution. Within this range, the optimal source size $R$ is obtained by minimizing the $\chi^2$, with $\lambda$ constrained to the experimentally determined interval of $0.64 \pm 0.06$. To explicitly account for the peak structure at higher momenta, corresponding to the $\Lambda(1520)$ resonance, an additional Breit-Wigner term is introduced~\cite{Kamiya:2019uiw,Liu:2026esv}
\begin{equation}
C_{\mathrm{res}}(k^*)=\frac{b\Gamma^{2}}{({k^*}^2/2\mu_{K^{-}p}+m_{K^{-}}+m_{p}-E_{R})^{2}+\Gamma^{2}/4}.
\end{equation}
The resonance parameters $E_{R}$ and $\Gamma$ extracted through this procedure are verified for consistency with the established mass and width of the $\Lambda(1520)$. We note that in Ref.~\cite{Liu:2026esv}, the $\Lambda(1520)$ is treated on the same footing as the $\Lambda(1405)$.

\begin{figure}[htbp]
    \centering
    \includegraphics[width=0.48\textwidth]{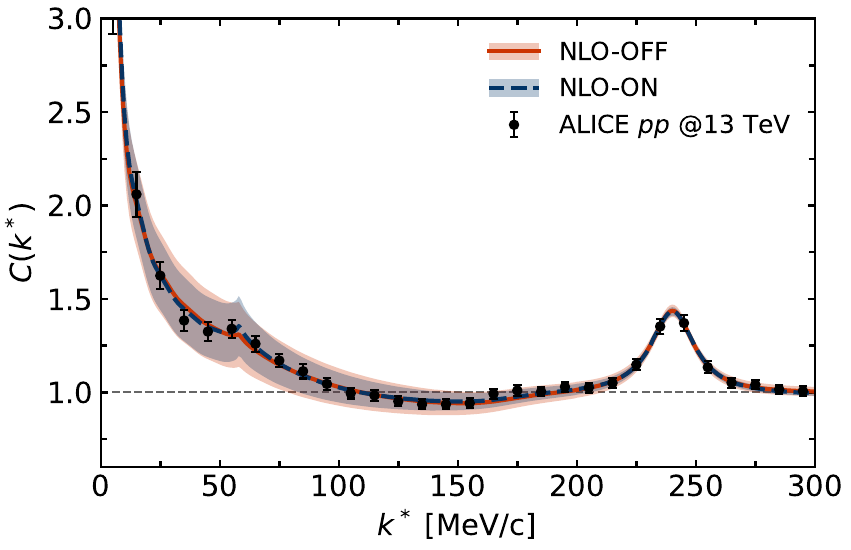}
    \caption{Comparison of the $K^-p$ correlation function $C(k^*)$ between the U$\chi$EFT fit results at NLO and ALICE $pp$ collision data at $\sqrt{s} = 13$ TeV~\cite{ALICE:2019gcn,ALICE:2022yyh}. The red solid line (NLO-OFF) represents the off-shell fit, while the blue dashed line (NLO-ON) corresponds to the on-shell approximation. The shaded bands indicate the uncertainties arising from variations in the ALICE channel weights~\cite{ALICE:2022yyh}.}
    \label{fig:CF_NLO_comparison}
\end{figure}

Using the standard channel weights $w_j$ from Ref.~\cite{ALICE:2022yyh}, the fit yields an unphysically small radius $R \approx 0.7$~fm ($R = 0.75 \pm 0.21$~fm at WT and $R = 0.83 \pm 0.27$~fm at NLO) resulting in a relatively larger $\chi^2 / \text{d.o.f.} \gtrsim 1.0$. 
This discrepancy arises because the conventional chiral model underestimates the $\bar{K}^0n$ coupling strength~\cite{ALICE:2022yyh}. To compensate for the resulting weak cusp, the fitting routine artificially compresses $R$ to sharpen the theoretical profile. By increasing $w_{\bar{K}^0n}$ by a factor of 2--3, consistent with the ALICE analysis, we find that $R$ naturally recovers to approximately 1~fm.
Using this optimized weighting ($w_{\bar{K}^0n} \approx 2.5$), the final NLO fits are presented in Fig.~\ref{fig:CF_NLO_comparison}.
\footnote{Ref.~\cite{Encarnacion:2024jge} argued that one can accommodate  the ALICE $K^-p$ CFs in the region of $k \lesssim 120$ MeV/c, considering theoretical uncertainties. As a result, this issue needs further study.}
Both the off-shell ($R = 1.06 \pm 0.11$~fm) and on-shell ($R = 0.99 \pm 0.20$~fm) schemes successfully describe the data. This degeneracy confirms that the uncertainties hinder the extraction of unique interaction parameters in both the source size and the channel weights.

The systematic differences between off-shell and on-shell CFs—which do not manifest as discernible experimental effects because they are offset by the adjustment of the source radius $R$—support the observation that the source function $S(\vec{r})$ is not completely independent of the underlying interaction~\cite{Epelbaum:2025aan,Molina:2025lzw}. Instead, $S(\vec{r})$ and the off-shell structure of the wave function are intertwined and defined only up to unitary transformations that leave on-shell observables invariant. This implies that assuming a universal Gaussian source without a consistent off-shell treatment may introduce certain model dependence in the extracted source size $R$~\cite{Wang:2024bpl,Xiong:2025bmd}.

\section{Summary and outlook} 
\label{sec:summary}
In this work, we have presented a systematic investigation of the meson-baryon SU(3) interactions in the $S=-1$ sector within unitarized chiral effective field theory. For the first time, an off-shell formulation has been implemented from leading order (LO) to next-to-leading order (NLO) using a finite cutoff regulator. This framework successfully resolves several long-standing conceptual limitations of traditional on-shell approaches, ensuring the rigorous restoration of proper analytic properties across the entire kinematic range.

Our results reveal a clear convergence pattern in the chiral expansion. While off-shell effects at the Weinberg-Tomozawa~(WT) level can be largely reabsorbed into the renormalization of fitted parameters, they become dynamically indispensable at NLO. At this order, the inclusion of the Born terms and higher-order contact interactions generates the essential momentum dependence required to describe high-precision data. The resulting dynamical consistency enables a robust determination of the two-pole structure of the $\Lambda(1405)$, reinforcing its modern interpretation as a prototypical dynamically generated resonance.

To scrutinize whether the off-shell dynamics yield a non-trivial contribution to the femtoscopic correlation functions (CFs), we decompose the contributions from each channel within the coupled-channel system, focusing specifically on the $K^-p$ channel. By employing the on-shell Koonin--Pratt (KP) formula with the scattering amplitude parametrized via the effective range expansion (ERE), the near-threshold behavior is predominantly characterized by the scattering length $a_0$. Our analysis reveals that the observed discrepancies between the NLO off-shell and on-shell results arise primarily from the different parameter sets obtained when fitting the experimental data, rather than from any substantial or directly observable effect of the off-shell dynamics themselves. 

However, extracting high-precision strong-interaction parameters remains challenging due to strong correlations among off-shell effects, the source size $R$, and the channel weights $w_j$. Our findings indicate that while an off-shell treatment provides a more formally consistent description, its practical advantage is currently limited by the uncertainties in the production environment~\cite{Molina:2025lzw}. Nonetheless, the off-shell formulation allows us to reliably predict the $\pi^\pm\Sigma^\mp$ correlation functions, which can play a nontrivial role in deciphering the nature of the $\Lambda(1405)$ and pertinent chiral dynamics.


\textit{Acknowledgment}.---This work is partly supported by the National Key R\&D Program of China under Grant No.~2023YFA1606703 and the National Natural Science Foundation of China under Grant No.~W2543006 and No.~12435007. J.X.L acknowledges support from the National Natural Science Foundation of China under Grant No.~12522505. Z.W.L acknowledges support from the National Natural Science Foundation of China under Grant No.~12405133 and No.~12347180, China Postdoctoral Science Foundation under Grant No.~2023M740189, and the Postdoctoral Fellowship Program of CPSF under Grant No.~GZC20233381. 

\bibliography{KbarN}

@article{Borasoy:2005fq,
    author = "Borasoy, B. and Nissler, R. and Weise, W.",
    title = "{Comment on `Surprises in threshold antikaon-nucleon physics'}",
    eprint = "hep-ph/0512279",
    archivePrefix = "arXiv",
    doi = "10.1103/PhysRevLett.96.199201",
    journal = "Phys. Rev. Lett.",
    volume = "96",
    pages = "199201",
    year = "2006"
}

@article{Borasoy:2004kk,
    author = "Borasoy, B. and Nissler, R. and Weise, W.",
    title = "{Kaonic hydrogen and K- p scattering}",
    eprint = "hep-ph/0410305",
    archivePrefix = "arXiv",
    doi = "10.1103/PhysRevLett.94.213401",
    journal = "Phys. Rev. Lett.",
    volume = "94",
    pages = "213401",
    year = "2005"
}

@article{Bayar:2011qj,
    author = "Bayar, M. and Yamagata-Sekihara, J. and Oset, E.",
    title = "{The $\bar{K}NN$ system with chiral dynamics}",
    eprint = "1102.2854",
    archivePrefix = "arXiv",
    primaryClass = "hep-ph",
    doi = "10.1103/PhysRevC.84.015209",
    journal = "Phys. Rev. C",
    volume = "84",
    pages = "015209",
    year = "2011"
}

@article{Ikeda:2010tk,
    author = "Ikeda, Yoichi and Kamano, Hiroyuki and Sato, Toru",
    title = "{Energy dependence of barKN interactions and resonance pole of strange dibaryons}",
    eprint = "1004.4877",
    archivePrefix = "arXiv",
    primaryClass = "nucl-th",
    reportNumber = "JLAB-THY-10-1172",
    doi = "10.1143/PTP.124.533",
    journal = "Prog. Theor. Phys.",
    volume = "124",
    pages = "533--539",
    year = "2010"
}

@article{Dote:2008hw,
    author = "Dote, Akinobu and Hyodo, Tetsuo and Weise, Wolfram",
    title = "{Variational calculation of the ppK- system based on chiral SU(3) dynamics}",
    eprint = "0806.4917",
    archivePrefix = "arXiv",
    primaryClass = "nucl-th",
    doi = "10.1103/PhysRevC.79.014003",
    journal = "Phys. Rev. C",
    volume = "79",
    pages = "014003",
    year = "2009"
}

@article{Dote:2008in,
    author = "Dote, Akinobu and Hyodo, Tetsuo and Weise, Wolfram",
    title = "{K- pp system with chiral SU(3) effective interaction}",
    eprint = "0802.0238",
    archivePrefix = "arXiv",
    primaryClass = "nucl-th",
    doi = "10.1016/j.nuclphysa.2008.02.001",
    journal = "Nucl. Phys. A",
    volume = "804",
    pages = "197--206",
    year = "2008"
}

@article{Morimatsu:2019wvk,
    author = "Morimatsu, Osamu and Yamada, Kazuki",
    title = "{Renormalization of the unitarized Weinberg-Tomozawa interaction without on-shell factorization and I=0 K{\textasciimacron}N{\textendash}{\ensuremath{\pi}}{\ensuremath{\Sigma}} coupled channels}",
    eprint = "1903.12380",
    archivePrefix = "arXiv",
    primaryClass = "hep-ph",
    reportNumber = "KEK-TH-2116",
    doi = "10.1103/PhysRevC.100.025201",
    journal = "Phys. Rev. C",
    volume = "100",
    number = "2",
    pages = "025201",
    year = "2019"
}

@article{Nieves:2001wt,
    author = "Nieves, J. and Ruiz Arriola, E.",
    title = "{The S(11) - N(1535) and - N(1650) resonances in meson baryon unitarized coupled channel chiral perturbation theory}",
    eprint = "hep-ph/0104307",
    archivePrefix = "arXiv",
    reportNumber = "UG-DFM-10-01",
    doi = "10.1103/PhysRevD.64.116008",
    journal = "Phys. Rev. D",
    volume = "64",
    pages = "116008",
    year = "2001"
}

@article{Revai:2017isg,
    author = "R{\'e}vai, J.",
    title = "{Are the chiral based $\bar{K}N$ potentials really energy dependent?}",
    eprint = "1711.04098",
    archivePrefix = "arXiv",
    primaryClass = "nucl-th",
    doi = "10.1007/s00601-018-1371-1",
    journal = "Few Body Syst.",
    volume = "59",
    number = "4",
    pages = "49",
    year = "2018"
}

@article{Ramos:2025ibe,
    author = "Ramos, {\`A}ngels and Torres-Rincon, Juan M. and de Fagoaga, Alejandro and Cabr{\'e}, Esteve",
    title = "{Kaon-deuteron femtoscopy from unitarized chiral interactions}",
    eprint = "2507.22593",
    archivePrefix = "arXiv",
    primaryClass = "hep-ph",
    doi = "10.1103/bhn4-m9nf",
    journal = "Phys. Rev. D",
    volume = "113",
    number = "3",
    pages = "036020",
    year = "2026"
}

@article{Nieves:2000km,
    author = "Nieves, J. and Ruiz Arriola, E.",
    title = "{Bethe-Salpeter approach for the P(33) elastic pion-nucleon scattering in heavy baryon chiral perturbation theory}",
    eprint = "hep-ph/0008034",
    archivePrefix = "arXiv",
    reportNumber = "UG-DFM-4-2000",
    doi = "10.1103/PhysRevD.63.076001",
    journal = "Phys. Rev. D",
    volume = "63",
    pages = "076001",
    year = "2001"
}

@article{He:2026mkf,
    author = "He, Ying-Bo and Liu, Xiao-Hai and Geng, Li-Sheng and Guo, Feng-Kun and Xie, Ju-Jun",
    title = "{Identifying the two-pole structure of the $\Lambda(1405)$ using an SU(3) flavor filter}",
    doi = "10.1103/5sr9-tzj6",
    journal = "Phys. Rev. D",
    volume = "113",
    number = "5",
    pages = "L051501",
    year = "2026"
}

@article{Nieves:1999bx,
    author = "Nieves, Juan and Ruiz Arriola, Enrique",
    title = "{Bethe-Salpeter approach for unitarized chiral perturbation theory}",
    eprint = "hep-ph/9907469",
    archivePrefix = "arXiv",
    reportNumber = "UG-DFM-2-99",
    doi = "10.1016/S0375-9474(00)00321-3",
    journal = "Nucl. Phys. A",
    volume = "679",
    pages = "57--117",
    year = "2000"
}

@article{Nieves:1998hp,
    author = "Nieves, Juan and Ruiz Arriola, Enrique",
    title = "{Bethe-Salpeter approach for meson meson scattering in chiral perturbation theory}",
    eprint = "nucl-th/9807035",
    archivePrefix = "arXiv",
    reportNumber = "UG-DFM-1-98",
    doi = "10.1016/S0370-2693(99)00461-X",
    journal = "Phys. Lett. B",
    volume = "455",
    pages = "30--38",
    year = "1999"
}

@article{Altenbuchinger:2013gaa,
    author = "Altenbuchinger, M. and Geng, Li-Sheng",
    title = "{Off-shell effects on the interaction of Nambu-Goldstone bosons and $D$ mesons}",
    eprint = "1310.5224",
    archivePrefix = "arXiv",
    primaryClass = "hep-ph",
    doi = "10.1103/PhysRevD.89.054008",
    journal = "Phys. Rev. D",
    volume = "89",
    number = "5",
    pages = "054008",
    year = "2014"
}

@inproceedings{Geng:2026ywt,
    author = "Geng, Li-Sheng and liu, Ming-Zhu and Xie, Jia-Ming",
    title = "{Strangeness is the key: from $\bar{K}N$ to $\bar{D}_s D K$}",
    booktitle = "{15th International Conference on Hypernuclear and Strange Particle Physics}",
    eprint = "2602.17190",
    archivePrefix = "arXiv",
    primaryClass = "hep-ph",
    month = "2",
    year = "2026"
}

@article{Xie:2025nnq,
    author = "Xie, Jia-Ming and Lu, Jun-Xu and Geng, Li-Sheng and Zou, Bing-Song",
    title = "{Two-pole structures in QCD-- a universal phenomenon governed by chiral dynamics}",
    eprint = "2504.03392",
    archivePrefix = "arXiv",
    primaryClass = "hep-ph",
    doi = "10.22323/1.479.0005",
    journal = "PoS",
    volume = "CD2024",
    pages = "005",
    year = "2026"
}

@article{Pittler:2025upn,
    author = "Pittler, Ferenc and Mai, Maxim and Mei{\ss}ner, Ulf-G. and Ferguson, Ryan F. and Hurck, Peter and Ireland, David G. and McKinnon, Bryan",
    title = "{Universal parameters of the {\ensuremath{\Lambda}}(1380), the {\ensuremath{\Lambda}}(1405), and their isospin partners from a combined analysis of lattice QCD and experimental results}",
    eprint = "2507.14283",
    archivePrefix = "arXiv",
    primaryClass = "hep-ph",
    doi = "10.1103/ls4c-6f2y",
    journal = "Phys. Rev. D",
    volume = "112",
    number = "7",
    pages = "074037",
    year = "2025"
}

@article{Liu:2026esv,
    author = "Liu, Si-Wei and Xie, Ju-Jun",
    title = "{Femtoscopic correlation functions for general partial waves: Application to the $Λ(1520)$ resonance}",
    eprint = "2601.22695",
    archivePrefix = "arXiv",
    primaryClass = "hep-ph",
    month = "1",
    year = "2026"
}

@article{Bethe:1949yr,
    author = "Bethe, H. A.",
    title = "{Theory of the Effective Range in Nuclear Scattering}",
    doi = "10.1103/PhysRev.76.38",
    journal = "Phys. Rev.",
    volume = "76",
    pages = "38--50",
    year = "1949"
}

@article{Ratcliffe:1998su,
    author = "Ratcliffe, Philip G.",
    title = "{SU(3) breaking in hyperon beta decays: A Prediction for xi0 ---{\ensuremath{>}} sigma+ e anti-neutrino}",
    eprint = "hep-ph/9806381",
    archivePrefix = "arXiv",
    reportNumber = "EPTCO-98-002",
    doi = "10.1103/PhysRevD.59.014038",
    journal = "Phys. Rev. D",
    volume = "59",
    pages = "014038",
    year = "1999"
}

@article{Pich:1995bw,
    author = "Pich, A.",
    title = "{Chiral perturbation theory}",
    eprint = "hep-ph/9502366",
    archivePrefix = "arXiv",
    reportNumber = "FTUV-95-4, IFIC-95-4",
    doi = "10.1088/0034-4885/58/6/001",
    journal = "Rept. Prog. Phys.",
    volume = "58",
    pages = "563--610",
    year = "1995"
}

@article{Li:2024tof,
    author = "Li, Hai-Peng and Yi, Jing-Yu and Xiao, Chu-Wen and Yao, De-Liang and Liang, Wei-Hong and Oset, Eulogio",
    title = "{Correlation function and the inverse problem in the BD interaction*}",
    eprint = "2401.14302",
    archivePrefix = "arXiv",
    primaryClass = "hep-ph",
    doi = "10.1088/1674-1137/ad2dc2",
    journal = "Chin. Phys. C",
    volume = "48",
    number = "5",
    pages = "053107",
    year = "2024"
}

@article{Khemchandani:2023xup,
    author = "Khemchandani, K. P. and Abreu, Luciano M. and Martinez Torres, A. and Navarra, F. S.",
    title = "{Can a femtoscopic correlation function shed light on the nature of the lightest charm axial mesons?}",
    eprint = "2312.11811",
    archivePrefix = "arXiv",
    primaryClass = "hep-ph",
    doi = "10.1103/PhysRevD.110.036008",
    journal = "Phys. Rev. D",
    volume = "110",
    number = "3",
    pages = "036008",
    year = "2024"
}

@article{Epelbaum:2014efa,
    author = "Epelbaum, E. and Krebs, H. and Mei{\ss}ner, U. G.",
    title = "{Improved chiral nucleon-nucleon potential up to next-to-next-to-next-to-leading order}",
    eprint = "1412.0142",
    archivePrefix = "arXiv",
    primaryClass = "nucl-th",
    doi = "10.1140/epja/i2015-15053-8",
    journal = "Eur. Phys. J. A",
    volume = "51",
    number = "5",
    pages = "53",
    year = "2015"
}

@article{Morita:2019rph,
    author = "Morita, Kenji and Gongyo, Shinya and Hatsuda, Tetsuo and Hyodo, Tetsuo and Kamiya, Yuki and Ohnishi, Akira",
    title = "{Probing $\Omega\Omega$ and $p\Omega$ dibaryons with femtoscopic correlations in relativistic heavy-ion collisions}",
    eprint = "1908.05414",
    archivePrefix = "arXiv",
    primaryClass = "nucl-th",
    reportNumber = "YITP-19-79, RIKEN-QHP-423, RIKEN-iTHEMS-Report-19",
    doi = "10.1103/PhysRevC.101.015201",
    journal = "Phys. Rev. C",
    volume = "101",
    number = "1",
    pages = "015201",
    year = "2020"
}

@article{Morita:2016auo,
    author = "Morita, Kenji and Ohnishi, Akira and Etminan, Faisal and Hatsuda, Tetsuo",
    title = "{Probing multistrange dibaryons with proton-omega correlations in high-energy heavy ion collisions}",
    eprint = "1605.06765",
    archivePrefix = "arXiv",
    primaryClass = "hep-ph",
    reportNumber = "YITP-16-62, RIKEN-QHP-223",
    doi = "10.1103/PhysRevC.94.031901",
    journal = "Phys. Rev. C",
    volume = "94",
    number = "3",
    pages = "031901",
    year = "2016",
    note = "[Erratum: Phys.Rev.C 100, 069902 (2019)]"
}

@article{CrystalBall:2001uhc,
    author = "Starostin, A. and others",
    collaboration = "Crystal Ball",
    title = "{Measurement of K- p --{\ensuremath{>}} eta Lambda near threshold}",
    doi = "10.1103/PhysRevC.64.055205",
    journal = "Phys. Rev. C",
    volume = "64",
    pages = "055205",
    year = "2001"
}

@article{Salpeter:1951sz,
    author = "Salpeter, E. E. and Bethe, H. A.",
    title = "{A Relativistic equation for bound state problems}",
    doi = "10.1103/PhysRev.84.1232",
    journal = "Phys. Rev.",
    volume = "84",
    pages = "1232--1242",
    year = "1951"
}

@article{ALICE:2019gcn,
    author = "Acharya, Shreyasi and others",
    collaboration = "ALICE",
    title = "{Scattering studies with low-energy kaon-proton femtoscopy in proton-proton collisions at the LHC}",
    eprint = "1905.13470",
    archivePrefix = "arXiv",
    primaryClass = "nucl-ex",
    reportNumber = "CERN-EP-2019-118",
    doi = "10.1103/PhysRevLett.124.092301",
    journal = "Phys. Rev. Lett.",
    volume = "124",
    number = "9",
    pages = "092301",
    year = "2020"
}

@article{ALICE:2022yyh,
    author = "Acharya, Shreyasi and others",
    collaboration = "ALICE",
    title = "{Constraining the ${\overline{\textrm{K}}}{\textrm{N}}$ coupled channel dynamics using femtoscopic correlations at the LHC}",
    eprint = "2205.15176",
    archivePrefix = "arXiv",
    primaryClass = "nucl-ex",
    reportNumber = "CERN-EP-2022-107",
    doi = "10.1140/epjc/s10052-023-11476-0",
    journal = "Eur. Phys. J. C",
    volume = "83",
    number = "4",
    pages = "340",
    year = "2023"
}

@article{Wang:2024bpl,
    author = "Wang, Lingxiao and Zhao, Jiaxing",
    title = "{Learning Hadron Emitting Sources with Deep Neural Networks}",
    eprint = "2411.16343",
    archivePrefix = "arXiv",
    primaryClass = "nucl-th",
    reportNumber = "RIKEN-iTHEMS-Report-24",
    month = "11",
    year = "2024"
}

@article{Xiong:2025bmd,
    author = "Xiong, Ao-Sheng and Yuan, Qi-Wei and Liu, Ming-Zhu and Yu, Fu-Sheng and Liu, Zhi-Wei and Geng, Li-Sheng",
    title = "{Solving the Inverse Source Problem in Femtoscopy with a Toy Model}",
    eprint = "2512.06904",
    archivePrefix = "arXiv",
    primaryClass = "hep-ph",
    month = "12",
    year = "2025"
}

@article{Liu:2024nac,
    author = "Liu, Zhi-Wei and Lu, Jun-Xu and Liu, Ming-Zhu and Geng, Li-Sheng",
    title = "{Femtoscopy can tell whether Zc(3900) and Zcs(3985) are resonances, virtual states, or bound states}",
    eprint = "2404.18607",
    archivePrefix = "arXiv",
    primaryClass = "hep-ph",
    doi = "10.1016/j.scib.2025.09.022",
    journal = "Sci. Bull.",
    volume = "70",
    pages = "3515--3521",
    year = "2025"
}

@article{Liu:2025oar,
    author = "Liu, Zhi-Wei and Ge, Duo-Lun and Lu, Jun-Xu and Liu, Ming-Zhu and Geng, Li-Sheng",
    title = "{Charmonium-nucleon femtoscopic correlation function}",
    eprint = "2504.04853",
    archivePrefix = "arXiv",
    primaryClass = "hep-ph",
    doi = "10.1103/3bdh-blwh",
    journal = "Phys. Rev. D",
    volume = "112",
    number = "5",
    pages = "054019",
    year = "2025"
}

@article{Mai:2022eur,
    author = "Mai, Maxim and Mei{\ss}ner, Ulf-G. and Urbach, Carsten",
    title = "{Towards a theory of hadron resonances}",
    eprint = "2206.01477",
    archivePrefix = "arXiv",
    primaryClass = "hep-ph",
    doi = "10.1016/j.physrep.2022.11.005",
    journal = "Phys. Rept.",
    volume = "1001",
    pages = "1--66",
    year = "2023"
}

@book{Weinberg:1996kr,
    author = "Weinberg, Steven",
    title = "{The quantum theory of fields. Vol. 2: Modern applications}",
    doi = "10.1017/CBO9781139644174",
    isbn = "978-1-139-63247-8, 978-0-521-67054-8, 978-0-521-55002-4",
    publisher = "Cambridge University Press",
    month = "8",
    year = "2013"
}

@article{Epelbaum:2025aan,
    author = "Epelbaum, Evgeny and Heihoff, Sven and Mei\ss{}ner, Ulf-G. and Tscherwon, Alexander",
    title = "{Can the strong interactions between hadrons be determined using femtoscopy?}",
    eprint = "2504.08631",
    archivePrefix = "arXiv",
    primaryClass = "nucl-th",
    month = "4",
    year = "2025"
}

@article{Capstick:1986ter,
    author = "Capstick, Simon and Isgur, Nathan",
    title = "{Baryons in a relativized quark model with chromodynamics}",
    doi = "10.1103/physrevd.34.2809",
    journal = "Phys. Rev. D",
    volume = "34",
    number = "9",
    pages = "2809--2835",
    year = "1986"
}

@article{Jennings:1986yg,
    author = "Jennings, B. K.",
    title = "{Further Evidence on the Nature of the $\Lambda$ (1405)}",
    reportNumber = "TRI-PP-86-32",
    doi = "10.1016/0370-2693(86)90955-X",
    journal = "Phys. Lett. B",
    volume = "176",
    pages = "229--232",
    year = "1986"
}

@article{Veit:1984jr,
    author = "Veit, E. A. and Jennings, Byron K. and Thomas, Anthony William and Barrett, R. C.",
    title = "{S Wave Meson - Nucleon Scattering in an SU(3) Cloudy Bag Model}",
    reportNumber = "TRI-PP-84-40, ADP-283",
    doi = "10.1103/PhysRevD.31.1033",
    journal = "Phys. Rev. D",
    volume = "31",
    pages = "1033",
    year = "1985"
}

@article{Veit:1984an,
    author = "Veit, E. A. and Jennings, Byron K. and Barrett, R. C. and Thomas, Anthony William",
    title = "{Kaon - Nucleon Scattering in an Extended Cloudy Bag Model}",
    reportNumber = "CERN-TH-3792, TRI-PP-83-123",
    doi = "10.1016/0370-2693(84)91746-5",
    journal = "Phys. Lett. B",
    volume = "137",
    pages = "415--418",
    year = "1984"
}

@article{Dalitz:1967fp,
    author = "Dalitz, R. H. and Wong, T. C. and Rajasekaran, G.",
    title = "{Model calculation for Y*(0) (1405) resonance state}",
    doi = "10.1103/PhysRev.153.1617",
    journal = "Phys. Rev.",
    volume = "153",
    pages = "1617--1623",
    year = "1967"
}

@article{Fink:1989uk,
    author = "Fink, Jr., P. J. and He, G. and Landau, R. H. and Schnick, J. W.",
    title = "{Bound States, Resonances and Poles in Low-energy $\bar{K} N$ Interaction Models}",
    reportNumber = "PRINT-89-0830 (OREGON-STATE)",
    doi = "10.1103/PhysRevC.41.2720",
    journal = "Phys. Rev. C",
    volume = "41",
    pages = "2720--2725",
    year = "1990"
}

@article{Tomozawa:1966jm,
    author = "Tomozawa, Y.",
    title = "{Axial vector coupling renormalization and the meson baryon scattering lengths}",
    doi = "10.1007/BF02857517",
    journal = "Nuovo Cim. A",
    volume = "46",
    pages = "707--717",
    year = "1966"
}

@article{Weinberg:1966kf,
    author = "Weinberg, Steven",
    title = "{Pion scattering lengths}",
    doi = "10.1103/PhysRevLett.17.616",
    journal = "Phys. Rev. Lett.",
    volume = "17",
    pages = "616--621",
    year = "1966"
}

@article{Sadasivan:2022srs,
    author = {Sadasivan, Daniel and Mai, Maxim and D\"oring, Michael and Mei\ss{}ner, Ulf-G. and Amorim, Felipe and Klucik, John Paul and Lu, Jun-Xu and Geng, Li-Sheng},
    title = "{New insights into the pole parameters of the $\Lambda(1380)$, the $\Lambda(1405)$ and the $\Sigma(1385)$}",
    eprint = "2212.10415",
    archivePrefix = "arXiv",
    primaryClass = "nucl-th",
    doi = "10.3389/fphy.2023.1139236",
    journal = "Front. Phys.",
    volume = "11",
    pages = "1139236",
    year = "2023"
}

@article{Oller:2005ig,
    author = "Oller, Jose A. and Prades, Joaquim and Verbeni, Michela",
    title = "{Surprises in threshold antikaon-nucleon physics}",
    eprint = "hep-ph/0508081",
    archivePrefix = "arXiv",
    reportNumber = "CAFPE-63-05, UG-FT-193-05",
    doi = "10.1103/PhysRevLett.95.172502",
    journal = "Phys. Rev. Lett.",
    volume = "95",
    pages = "172502",
    year = "2005"
}

@article{BaryonScatteringBaSc:2023zvt,
    author = "Bulava, John and others",
    collaboration = "Baryon Scattering (BaSc)",
    title = "{Two-Pole Nature of the \ensuremath{\Lambda}(1405) resonance from Lattice QCD}",
    eprint = "2307.10413",
    archivePrefix = "arXiv",
    primaryClass = "hep-lat",
    reportNumber = "MIT-CTP/5579",
    doi = "10.1103/PhysRevLett.132.051901",
    journal = "Phys. Rev. Lett.",
    volume = "132",
    number = "5",
    pages = "051901",
    year = "2024"
}

@article{BaryonScatteringBaSc:2023ori,
    author = "Bulava, John and others",
    collaboration = "Baryon Scattering (BaSc)",
    title = "{Lattice QCD study of \ensuremath{\pi}\ensuremath{\Sigma}-K\textasciimacron{}N scattering and the \ensuremath{\Lambda}(1405) resonance}",
    eprint = "2307.13471",
    archivePrefix = "arXiv",
    primaryClass = "hep-lat",
    reportNumber = "MIT-CTP/5580",
    doi = "10.1103/PhysRevD.109.014511",
    journal = "Phys. Rev. D",
    volume = "109",
    number = "1",
    pages = "014511",
    year = "2024"
}

@article{Weinberg:1990rz,
    author = "Weinberg, Steven",
    title = "{Nuclear forces from chiral Lagrangians}",
    reportNumber = "UTTG-31-90",
    doi = "10.1016/0370-2693(90)90938-3",
    journal = "Phys. Lett. B",
    volume = "251",
    pages = "288--292",
    year = "1990"
}

@article{Oller:1997ng,
    author = "Oller, J. A. and Oset, E. and Pelaez, J. R.",
    title = "{Nonperturbative approach to effective chiral Lagrangians and meson interactions}",
    eprint = "hep-ph/9803242",
    archivePrefix = "arXiv",
    doi = "10.1103/PhysRevLett.80.3452",
    journal = "Phys. Rev. Lett.",
    volume = "80",
    pages = "3452--3455",
    year = "1998"
}

@article{Oller:1998zr,
    author = "Oller, J. A. and Oset, E.",
    title = "{N/D description of two meson amplitudes and chiral symmetry}",
    eprint = "hep-ph/9809337",
    archivePrefix = "arXiv",
    doi = "10.1103/PhysRevD.60.074023",
    journal = "Phys. Rev. D",
    volume = "60",
    pages = "074023",
    year = "1999"
}

@article{Oller:1998hw,
    author = "Oller, J. A. and Oset, E. and Pelaez, J. R.",
    title = "{Meson meson interaction in a nonperturbative chiral approach}",
    eprint = "hep-ph/9804209",
    archivePrefix = "arXiv",
    reportNumber = "SLAC-PUB-7787",
    doi = "10.1103/PhysRevD.59.074001",
    journal = "Phys. Rev. D",
    volume = "59",
    pages = "074001",
    year = "1999",
    note = "[Erratum: Phys.Rev.D 60, 099906 (1999), Erratum: Phys.Rev.D 75, 099903 (2007)]"
}

@article{Gasser:1984gg,
    author = "Gasser, J. and Leutwyler, H.",
    title = "{Chiral Perturbation Theory: Expansions in the Mass of the Strange Quark}",
    reportNumber = "CERN-TH-3798",
    doi = "10.1016/0550-3213(85)90492-4",
    journal = "Nucl. Phys. B",
    volume = "250",
    pages = "465--516",
    year = "1985"
}

@article{Gasser:1983yg,
    author = "Gasser, J. and Leutwyler, H.",
    title = "{Chiral Perturbation Theory to One Loop}",
    reportNumber = "CERN-TH-3689",
    doi = "10.1016/0003-4916(84)90242-2",
    journal = "Annals Phys.",
    volume = "158",
    pages = "142",
    year = "1984"
}

@article{Weinberg:1978kz,
    author = "Weinberg, Steven",
    editor = "Deser, S.",
    title = "{Phenomenological Lagrangians}",
    reportNumber = "HUTP-78-A051A",
    doi = "10.1016/0378-4371(79)90223-1",
    journal = "Physica A",
    volume = "96",
    number = "1-2",
    pages = "327--340",
    year = "1979"
}

@article{Machleidt:2011zz,
    author = "Machleidt, R. and Entem, D. R.",
    title = "{Chiral effective field theory and nuclear forces}",
    eprint = "1105.2919",
    archivePrefix = "arXiv",
    primaryClass = "nucl-th",
    doi = "10.1016/j.physrep.2011.02.001",
    journal = "Phys. Rept.",
    volume = "503",
    pages = "1--75",
    year = "2011"
}

@article{Liu:2024uxn,
    author = "Liu, Ming-Zhu and Pan, Ya-Wen and Liu, Zhi-Wei and Wu, Tian-Wei and Lu, Jun-Xu and Geng, Li-Sheng",
    title = "{Three ways to decipher the nature of exotic hadrons: Multiplets, three-body hadronic molecules, and correlation functions}",
    eprint = "2404.06399",
    archivePrefix = "arXiv",
    primaryClass = "hep-ph",
    doi = "10.1016/j.physrep.2024.12.001",
    journal = "Phys. Rept.",
    volume = "1108",
    pages = "1--108",
    year = "2025"
}

@article{Oller:1997ti,
    author = "Oller, J. A. and Oset, E.",
    title = "{Chiral symmetry amplitudes in the S wave isoscalar and isovector channels and the $\sigma$, f$_0$(980), a$_0$(980) scalar mesons}",
    eprint = "hep-ph/9702314",
    archivePrefix = "arXiv",
    doi = "10.1016/S0375-9474(97)00160-7",
    journal = "Nucl. Phys. A",
    volume = "620",
    pages = "438--456",
    year = "1997",
    note = "[Erratum: Nucl.Phys.A 652, 407--409 (1999)]"
}

@article{Molina:2025lzw,
    author = "Molina, R. and Oset, E.",
    title = "{Determination of off-shell ambiguities in correlation functions: Strategies to minimize them}",
    eprint = "2506.03669",
    archivePrefix = "arXiv",
    primaryClass = "hep-ph",
    doi = "10.1103/rst4-rkmm",
    journal = "Phys. Rev. D",
    volume = "112",
    number = "9",
    pages = "096006",
    year = "2025"
}

@article{Zhuang:2024udv,
    author = "Zhuang, Zejian and Molina, Raquel and Lu, Jun-Xu and Geng, Li-Sheng",
    title = "{Pole trajectories of the {\ensuremath{\Lambda}}(1405) help establish its dynamical nature}",
    eprint = "2405.07686",
    archivePrefix = "arXiv",
    primaryClass = "hep-ph",
    doi = "10.1016/j.scib.2025.04.029",
    journal = "Sci. Bull.",
    volume = "70",
    pages = "1953--1961",
    year = "2025"
}

@article{Guo:2023wes,
    author = "Guo, Feng-Kun and Kamiya, Yuki and Mai, Maxim and Mei\ss{}ner, Ulf-G.",
    title = "{New insights into the nature of the \ensuremath{\Lambda}(1380) and \ensuremath{\Lambda}(1405) resonances away from the SU(3) limit}",
    eprint = "2308.07658",
    archivePrefix = "arXiv",
    primaryClass = "hep-ph",
    doi = "10.1016/j.physletb.2023.138264",
    journal = "Phys. Lett. B",
    volume = "846",
    pages = "138264",
    year = "2023"
}

@article{Xie:2023jve,
    author = "Xie, Jia-Ming and Lu, Jun-Xu and Geng, Li-Sheng and Zou, Bing-Song",
    title = "{Dynamical origin of universal two-pole structures and their light quark mass evolution}",
    eprint = "2312.17287",
    archivePrefix = "arXiv",
    primaryClass = "hep-ph",
    doi = "10.1051/epjconf/202430301011",
    journal = "EPJ Web Conf.",
    volume = "303",
    pages = "01011",
    year = "2024"
}

@article{Xie:2023cej,
    author = "Xie, Jia-Ming and Lu, Jun-Xu and Geng, Li-Sheng and Zou, Bing-Song",
    title = "{Two-pole structures as a universal phenomenon dictated by coupled-channel chiral dynamics}",
    eprint = "2307.11631",
    archivePrefix = "arXiv",
    primaryClass = "hep-ph",
    doi = "10.1103/PhysRevD.108.L111502",
    journal = "Phys. Rev. D",
    volume = "108",
    number = "11",
    pages = "L111502",
    year = "2023"
}

@article{Oller:2000ma,
    author = "Oller, J. A. and Oset, E. and Ramos, A.",
    title = "{Chiral unitary approach to meson meson and meson - baryon interactions and nuclear applications}",
    eprint = "hep-ph/0002193",
    archivePrefix = "arXiv",
    reportNumber = "FZJ-IKP-TH-1999-37, FTUV-99-1215, IFIC-99-1215",
    doi = "10.1016/S0146-6410(00)00104-6",
    journal = "Prog. Part. Nucl. Phys.",
    volume = "45",
    pages = "157--242",
    year = "2000"
}

@article{Oller:2019opk,
    author = "Oller, J. A.",
    title = "{Coupled-channel approach in hadron\textendash{}hadron scattering}",
    eprint = "1909.00370",
    archivePrefix = "arXiv",
    primaryClass = "hep-ph",
    doi = "10.1016/j.ppnp.2019.103728",
    journal = "Prog. Part. Nucl. Phys.",
    volume = "110",
    pages = "103728",
    year = "2020"
}

@article{Kaiser:1996js,
    author = "Kaiser, Norbert and Waas, T. and Weise, W.",
    title = "{SU(3) chiral dynamics with coupled channels: Eta and kaon photoproduction}",
    eprint = "hep-ph/9607459",
    archivePrefix = "arXiv",
    doi = "10.1016/S0375-9474(96)00321-1",
    journal = "Nucl. Phys. A",
    volume = "612",
    pages = "297--320",
    year = "1997"
}

@article{Callan:1985hy,
    author = "Callan, Jr., Curtis G. and Klebanov, Igor R.",
    title = "{Bound State Approach to Strangeness in the Skyrme Model}",
    reportNumber = "Print-85-0733 (PRINCETON)",
    doi = "10.1016/0550-3213(85)90292-5",
    journal = "Nucl. Phys. B",
    volume = "262",
    pages = "365--382",
    year = "1985"
}

@article{Klempt:2009pi,
    author = "Klempt, Eberhard and Richard, Jean-Marc",
    title = "{Baryon spectroscopy}",
    eprint = "0901.2055",
    archivePrefix = "arXiv",
    primaryClass = "hep-ph",
    doi = "10.1103/RevModPhys.82.1095",
    journal = "Rev. Mod. Phys.",
    volume = "82",
    pages = "1095--1153",
    year = "2010"
}

@article{Curceanu:2019uph,
    author = "Curceanu, Catalina and others",
    title = "{The modern era of light kaonic atom experiments}",
    doi = "10.1103/RevModPhys.91.025006",
    journal = "Rev. Mod. Phys.",
    volume = "91",
    number = "2",
    pages = "025006",
    year = "2019"
}

@article{Hyodo:2011ur,
    author = "Hyodo, Tetsuo and Jido, Daisuke",
    title = "{The nature of the Lambda(1405) resonance in chiral dynamics}",
    eprint = "1104.4474",
    archivePrefix = "arXiv",
    primaryClass = "nucl-th",
    doi = "10.1016/j.ppnp.2011.07.002",
    journal = "Prog. Part. Nucl. Phys.",
    volume = "67",
    pages = "55--98",
    year = "2012"
}

@article{Politzer:1973fx,
    author = "Politzer, H. David",
    editor = "Taylor, J. C.",
    title = "{Reliable Perturbative Results for Strong Interactions?}",
    doi = "10.1103/PhysRevLett.30.1346",
    journal = "Phys. Rev. Lett.",
    volume = "30",
    pages = "1346--1349",
    year = "1973"
}

@article{Gross:1973id,
    author = "Gross, David J. and Wilczek, Frank",
    editor = "Taylor, J. C.",
    title = "{Ultraviolet Behavior of Nonabelian Gauge Theories}",
    doi = "10.1103/PhysRevLett.30.1343",
    journal = "Phys. Rev. Lett.",
    volume = "30",
    pages = "1343--1346",
    year = "1973"
}

@article{Gross:1973ju,
    author = "Gross, D. J. and Wilczek, Frank",
    title = "{Asymptotically Free Gauge Theories - I}",
    reportNumber = "NAL-PUB-73-49-THY, FERMILAB-PUB-73-049-T",
    doi = "10.1103/PhysRevD.8.3633",
    journal = "Phys. Rev. D",
    volume = "8",
    pages = "3633--3652",
    year = "1973"
}

@article{Gross:2022hyw,
    author = "Gross, Franz and others",
    title = "{50 Years of Quantum Chromodynamics}",
    eprint = "2212.11107",
    archivePrefix = "arXiv",
    primaryClass = "hep-ph",
    doi = "10.1140/epjc/s10052-023-11949-2",
    journal = "Eur. Phys. J. C",
    volume = "83",
    pages = "1125",
    year = "2023"
}

@article{Lu:2022hwm,
    author = "Lu, Jun-Xu and Geng, Li-Sheng and Doering, Michael and Mai, Maxim",
    title = "{Cross-Channel Constraints on Resonant Antikaon-Nucleon Scattering}",
    eprint = "2209.02471",
    archivePrefix = "arXiv",
    primaryClass = "hep-ph",
    doi = "10.1103/PhysRevLett.130.071902",
    journal = "Phys. Rev. Lett.",
    volume = "130",
    number = "7",
    pages = "071902",
    year = "2023"
}

@article{Oller:2006yh,
    author = "Oller, Jose Antonio and Verbeni, Michela and Prades, Joaquim",
    title = "{Meson-baryon effective chiral lagrangians to O(q**3)}",
    eprint = "hep-ph/0608204",
    archivePrefix = "arXiv",
    reportNumber = "CAFPE-69-06, UGFT-199-06",
    doi = "10.1088/1126-6708/2006/09/079",
    journal = "JHEP",
    volume = "09",
    pages = "079",
    year = "2006"
}

@article{Gal:2016boi,
    author = "Gal, A. and Hungerford, E. V. and Millener, D. J.",
    title = "{Strangeness in nuclear physics}",
    eprint = "1605.00557",
    archivePrefix = "arXiv",
    primaryClass = "nucl-th",
    doi = "10.1103/RevModPhys.88.035004",
    journal = "Rev. Mod. Phys.",
    volume = "88",
    number = "3",
    pages = "035004",
    year = "2016"
}

@article{Batty:1997zp,
    author = "Batty, C. J. and Friedman, E. and Gal, A.",
    title = "{Strong interaction physics from hadronic atoms}",
    doi = "10.1016/S0370-1573(97)00011-2",
    journal = "Phys. Rept.",
    volume = "287",
    pages = "385--445",
    year = "1997"
}

@article{Kaiser:1995eg,
    author = "Kaiser, Norbert and Siegel, P. B. and Weise, W.",
    title = "{Chiral dynamics and the low-energy kaon - nucleon interaction}",
    eprint = "nucl-th/9505043",
    archivePrefix = "arXiv",
    reportNumber = "TUM-T39-95-5",
    doi = "10.1016/0375-9474(95)00362-5",
    journal = "Nucl. Phys. A",
    volume = "594",
    pages = "325--345",
    year = "1995"
}

@article{Oset:1997it,
    author = "Oset, E. and Ramos, A.",
    title = "{Nonperturbative chiral approach to s wave anti-K N interactions}",
    eprint = "nucl-th/9711022",
    archivePrefix = "arXiv",
    doi = "10.1016/S0375-9474(98)00170-5",
    journal = "Nucl. Phys. A",
    volume = "635",
    pages = "99--120",
    year = "1998"
}

@article{Oller:2000fj,
    author = "Oller, J. A. and Mei{\ss}ner, Ulf-G.",
    title = "{Chiral dynamics in the presence of bound states: Kaon nucleon interactions revisited}",
    eprint = "hep-ph/0011146",
    archivePrefix = "arXiv",
    reportNumber = "FZJ-IKP-TH-2000-26",
    doi = "10.1016/S0370-2693(01)00078-8",
    journal = "Phys. Lett. B",
    volume = "500",
    pages = "263--272",
    year = "2001"
}

@article{Borasoy:2006sr,
    author = "Borasoy, B. and Mei{\ss}ner, Ulf-G. and Nissler, R.",
    title = "{K- p scattering length from scattering experiments}",
    eprint = "hep-ph/0606108",
    archivePrefix = "arXiv",
    reportNumber = "HISKP-TH-06-15, FZJ-IKP-TH-2006-16",
    doi = "10.1103/PhysRevC.74.055201",
    journal = "Phys. Rev. C",
    volume = "74",
    pages = "055201",
    year = "2006"
}

@article{Ikeda:2012au,
    author = "Ikeda, Yoichi and Hyodo, Tetsuo and Weise, Wolfram",
    title = "{Chiral SU(3) theory of antikaon-nucleon interactions with improved threshold constraints}",
    eprint = "1201.6549",
    archivePrefix = "arXiv",
    primaryClass = "nucl-th",
    doi = "10.1016/j.nuclphysa.2012.01.029",
    journal = "Nucl. Phys. A",
    volume = "881",
    pages = "98--114",
    year = "2012"
}

@article{Guo:2012vv,
    author = "Guo, Zhi-Hui and Oller, J. A.",
    title = "{Meson-baryon reactions with strangeness -1 within a chiral framework}",
    eprint = "1210.3485",
    archivePrefix = "arXiv",
    primaryClass = "hep-ph",
    doi = "10.1103/PhysRevC.87.035202",
    journal = "Phys. Rev. C",
    volume = "87",
    number = "3",
    pages = "035202",
    year = "2013"
}

@article{Mai:2012dt,
    author = "Mai, Maxim and Mei{\ss}ner, Ulf-G.",
    title = "{New insights into antikaon-nucleon scattering and the structure of the Lambda(1405)}",
    eprint = "1202.2030",
    archivePrefix = "arXiv",
    primaryClass = "nucl-th",
    doi = "10.1016/j.nuclphysa.2013.01.032",
    journal = "Nucl. Phys. A",
    volume = "900",
    pages = "51 -- 64",
    year = "2013"
}

@article{Ramos:2016odk,
    author = "Ramos, A. and Feijoo, A. and Magas, V. K.",
    title = "{The chiral S = \ensuremath{-}1 meson\textendash{}baryon interaction with new constraints on the NLO contributions}",
    eprint = "1605.03767",
    archivePrefix = "arXiv",
    primaryClass = "nucl-th",
    doi = "10.1016/j.nuclphysa.2016.05.006",
    journal = "Nucl. Phys. A",
    volume = "954",
    pages = "58--74",
    year = "2016"
}

@article{Meissner:2020khl,
    author = "Mei\ss{}ner, Ulf-G.",
    title = "{Two-pole structures in QCD: Facts, not fantasy!}",
    eprint = "2005.06909",
    archivePrefix = "arXiv",
    primaryClass = "hep-ph",
    doi = "10.3390/sym12060981",
    journal = "Symmetry",
    volume = "12",
    number = "6",
    pages = "981",
    year = "2020"
}

@article{Mai:2020ltx,
    author = "Mai, Maxim",
    title = "{Review of the ${\Lambda }$(1405) A curious case of a strangeness resonance}",
    eprint = "2010.00056",
    archivePrefix = "arXiv",
    primaryClass = "nucl-th",
    doi = "10.1140/epjs/s11734-021-00144-7",
    journal = "Eur. Phys. J. ST",
    volume = "230",
    number = "6",
    pages = "1593--1607",
    year = "2021"
}

@article{Hyodo:2020czb,
    author = "Hyodo, Tetsuo and Niiyama, Masayuki",
    title = "{QCD and the strange baryon spectrum}",
    eprint = "2010.07592",
    archivePrefix = "arXiv",
    primaryClass = "hep-ph",
    doi = "10.1016/j.ppnp.2021.103868",
    journal = "Prog. Part. Nucl. Phys.",
    volume = "120",
    pages = "103868",
    year = "2021"
}

@article{Humphrey:1962zz,
    author = "Humphrey, William E. and Ross, Ronald R.",
    title = "{Low-Energy Interactions of K- Mesons in Hydrogen}",
    doi = "10.1103/PhysRev.127.1305",
    journal = "Phys. Rev.",
    volume = "127",
    pages = "1305--1323",
    year = "1962"
}

@article{Kim:1965zzd,
    author = "Kim, J. K.",
    title = "{Low Energy K- p Interaction of the 1405 MeV Y\textasciicircum{}*()0 Resonance as KbarN Bound State}",
    doi = "10.1103/PhysRevLett.14.29",
    journal = "Phys. Rev. Lett.",
    volume = "14",
    pages = "29",
    year = "1965"
}

@article{Sakitt:1965kh,
    author = "Sakitt, M. and Day, T. B. and Glasser, R. G. and Seeman, N. and Friedman, J. H. and Humphrey, W. E. and Ross, R. R.",
    title = "{LOW-ENERGY K- MESON INTERACTIONS IN HYDROGEN}",
    doi = "10.1103/PhysRev.139.B719",
    journal = "Phys. Rev.",
    volume = "139",
    pages = "B719",
    year = "1965"
}

@article{Kittel:1966zz,
    author = "Kittel, W. and Otter, G. and Wacek, I.",
    title = "{The K- p Charge Exchange Interactions at Low Energies and Scattering Lengths Determination}",
    doi = "10.1016/0031-9163(66)90845-6",
    journal = "Phys. Lett.",
    volume = "21",
    pages = "349--351",
    year = "1966"
}

@article{Evans:1983hz,
    author = "Evans, D. and Major, J. V. and Rondio, E. and Zakrzewski, Janusz Andrzej and Conboy, J. E. and Miller, D. J. and Tymieniecka, T.",
    title = "{CHARGE EXCHANGE SCATTERING IN K- P INTERACTIONS BELOW 300-MEV/C}",
    doi = "10.1088/0305-4616/9/8/011",
    journal = "J. Phys. G",
    volume = "9",
    pages = "885--894",
    year = "1983"
}

@article{Ciborowski:1982et,
    author = "Ciborowski, J. and others",
    title = "{KAON SCATTERING AND CHARGED SIGMA HYPERON PRODUCTION IN K- P INTERACTIONS BELOW 300-MEV/C}",
    doi = "10.1088/0305-4616/8/1/005",
    journal = "J. Phys. G",
    volume = "8",
    pages = "13--32",
    year = "1982"
}

@article{Nowak:1978au,
    author = "Nowak, R. J. and others",
    title = "{Charged Sigma Hyperon Production by K- Meson Interactions at Rest}",
    doi = "10.1016/0550-3213(78)90179-7",
    journal = "Nucl. Phys. B",
    volume = "139",
    pages = "61--71",
    year = "1978"
}

@article{Tovee:1971ga,
    author = "Tovee, D. N. and others",
    title = "{Some properties of the charged sigma hyperons}",
    doi = "10.1016/0550-3213(71)90302-6",
    journal = "Nucl. Phys. B",
    volume = "33",
    pages = "493--504",
    year = "1971"
}

@article{Bazzi:2011zj,
    author = "Bazzi, M. and others",
    collaboration = "SIDDHARTA",
    title = "{A New Measurement of Kaonic Hydrogen X-rays}",
    eprint = "1105.3090",
    archivePrefix = "arXiv",
    primaryClass = "nucl-ex",
    doi = "10.1016/j.physletb.2011.09.011",
    journal = "Phys. Lett. B",
    volume = "704",
    pages = "113--117",
    year = "2011"
}

@article{Meissner:2004jr,
    author = "Mei{\ss}ner, Ulf-G. and Raha, Udit and Rusetsky, Akaki",
    title = "{Spectrum and decays of kaonic hydrogen}",
    eprint = "hep-ph/0402261",
    archivePrefix = "arXiv",
    reportNumber = "HISKP-TH-04-01",
    doi = "10.1140/epjc/s2004-01859-4",
    journal = "Eur. Phys. J. C",
    volume = "35",
    pages = "349--357",
    year = "2004"
}

@article{Furnstahl:2015rha,
    author = "Furnstahl, R. J. and Klco, N. and Phillips, D. R. and Wesolowski, S.",
    title = "{Quantifying truncation errors in effective field theory}",
    eprint = "1506.01343",
    archivePrefix = "arXiv",
    primaryClass = "nucl-th",
    doi = "10.1103/PhysRevC.92.024005",
    journal = "Phys. Rev. C",
    volume = "92",
    number = "2",
    pages = "024005",
    year = "2015"
}

@article{Melendez:2017phj,
    author = "Melendez, J. A. and Wesolowski, S. and Furnstahl, R. J.",
    title = "{Bayesian truncation errors in chiral effective field theory: nucleon-nucleon observables}",
    eprint = "1704.03308",
    archivePrefix = "arXiv",
    primaryClass = "nucl-th",
    doi = "10.1103/PhysRevC.96.024003",
    journal = "Phys. Rev. C",
    volume = "96",
    number = "2",
    pages = "024003",
    year = "2017"
}

@article{Melendez:2019izc,
    author = "Melendez, J. A. and Furnstahl, R. J. and Phillips, D. R. and Pratola, M. T. and Wesolowski, S.",
    title = "{Quantifying Correlated Truncation Errors in Effective Field Theory}",
    eprint = "1904.10581",
    archivePrefix = "arXiv",
    primaryClass = "nucl-th",
    doi = "10.1103/PhysRevC.100.044001",
    journal = "Phys. Rev. C",
    volume = "100",
    number = "4",
    pages = "044001",
    year = "2019"
}

@article{Dalitz:1959dn,
    author = "Dalitz, R. H. and Tuan, S. F.",
    title = "{A possible resonant state in pion-hyperon scattering}",
    doi = "10.1103/PhysRevLett.2.425",
    journal = "Phys. Rev. Lett.",
    volume = "2",
    pages = "425--428",
    year = "1959"
}

@article{Jido:2003cb,
    author = "Jido, D. and Oller, J. A. and Oset, E. and Ramos, A. and Mei{\ss}ner, Ulf-G.",
    title = "{Chiral dynamics of the two Lambda(1405) states}",
    eprint = "nucl-th/0303062",
    archivePrefix = "arXiv",
    doi = "10.1016/S0375-9474(03)01598-7",
    journal = "Nucl. Phys. A",
    volume = "725",
    pages = "181--200",
    year = "2003"
}

@article{Ikeda:2011pi,
    author = "Ikeda, Yoichi and Hyodo, Tetsuo and Weise, Wolfram",
    title = "{Improved constraints on chiral SU(3) dynamics from kaonic hydrogen}",
    eprint = "1109.3005",
    archivePrefix = "arXiv",
    primaryClass = "nucl-th",
    doi = "10.1016/j.physletb.2011.10.068",
    journal = "Phys. Lett. B",
    volume = "706",
    pages = "63--67",
    year = "2011"
}

@article{Sadasivan:2018jig,
    author = {Sadasivan, D. and Mai, M. and D\"oring, M.},
    title = "{S- and p-wave structure of $S=-1$ meson-baryon scattering in the resonance region}",
    eprint = "1805.04534",
    archivePrefix = "arXiv",
    primaryClass = "nucl-th",
    reportNumber = "JLAB-THY-18-2699",
    doi = "10.1016/j.physletb.2018.12.035",
    journal = "Phys. Lett. B",
    volume = "789",
    pages = "329--335",
    year = "2019"
}

@article{Epelbaum:2019zqc,
    author = "Epelbaum, E. and others",
    title = "{Towards high-order calculations of three-nucleon scattering in chiral effective field theory}",
    eprint = "1907.03608",
    archivePrefix = "arXiv",
    primaryClass = "nucl-th",
    doi = "10.1140/epja/s10050-020-00102-2",
    journal = "Eur. Phys. J. A",
    volume = "56",
    number = "3",
    pages = "92",
    year = "2020"
}

@article{Lu:2021gsb,
    author = "Lu, Jun-Xu and Wang, Chun-Xuan and Xiao, Yang and Geng, Li-Sheng and Meng, Jie and Ring, Peter",
    title = "{Accurate Relativistic Chiral Nucleon-Nucleon Interaction up to Next-to-Next-to-Leading Order}",
    eprint = "2111.07766",
    archivePrefix = "arXiv",
    primaryClass = "nucl-th",
    doi = "10.1103/PhysRevLett.128.142002",
    journal = "Phys. Rev. Lett.",
    volume = "128",
    number = "14",
    pages = "142002",
    year = "2022"
}

@article{Borasoy:2005ie,
    author = "Borasoy, B. and Nissler, R. and Weise, W.",
    title = "{Chiral dynamics of kaon-nucleon interactions, revisited}",
    eprint = "hep-ph/0505239",
    archivePrefix = "arXiv",
    doi = "10.1140/epja/i2005-10079-1",
    journal = "Eur. Phys. J. A",
    volume = "25",
    pages = "79--96",
    year = "2005"
}

@article{Liu:2025rci,
    author = "Liu, Zhi-Wei and Lu, Jun-Xu and Geng, Li-Sheng",
    title = "{The femtoscopic technique{\textemdash}an invaluable tool in studies of exotic hadrons}",
    doi = "10.22323/1.465.0044",
    journal = "PoS",
    volume = "QNP2024",
    pages = "044",
    year = "2025"
}

@article{Shen:2025qpj,
    author = "Shen, Yi-bo and Liu, Zhi-Wei and Lu, Jun-Xu and Liu, Ming-Zhu and Geng, Li-Sheng",
    title = "{Probing the structure of exotic hadrons through correlation functions}",
    eprint = "2506.23476",
    archivePrefix = "arXiv",
    primaryClass = "hep-ph",
    month = "6",
    year = "2025"
}

@article{Koonin:1977fh,
    author = "Koonin, S. E.",
    title = "{Proton Pictures of High-Energy Nuclear Collisions}",
    doi = "10.1016/0370-2693(77)90340-9",
    journal = "Phys. Lett. B",
    volume = "70",
    pages = "43--47",
    year = "1977"
}

@article{Pratt:1990zq,
    author = "Pratt, S. and Csorgo, T. and Zimanyi, J.",
    title = "{Detailed predictions for two pion correlations in ultrarelativistic heavy ion collisions}",
    doi = "10.1103/PhysRevC.42.2646",
    journal = "Phys. Rev. C",
    volume = "42",
    pages = "2646--2652",
    year = "1990"
}

@article{Encarnacion:2024jge,
    author = "Encarnaci{\'o}n, P. and Feijoo, A. and Sarti, V. Mantovani and Ramos, A.",
    title = "{Femtoscopic study of the S=-1 meson-baryon interaction: K-p, {\ensuremath{\pi}}-{\ensuremath{\Lambda}}, and K+{\ensuremath{\Xi}}- correlations}",
    eprint = "2412.20880",
    archivePrefix = "arXiv",
    primaryClass = "hep-ph",
    doi = "10.1103/3ycr-vzmd",
    journal = "Phys. Rev. D",
    volume = "111",
    number = "11",
    pages = "114013",
    year = "2025"
}

@article{Feijoo:2024bvn,
    author = "Feijoo, A. and Korwieser, M. and Fabbietti, L.",
    title = "{Relevance of the coupled channels in the {\ensuremath{\phi}}p and {\ensuremath{\rho}}0p correlation functions}",
    eprint = "2407.01128",
    archivePrefix = "arXiv",
    primaryClass = "hep-ph",
    doi = "10.1103/PhysRevD.111.014009",
    journal = "Phys. Rev. D",
    volume = "111",
    number = "1",
    pages = "014009",
    year = "2025"
}

@article{Sarti:2023wlg,
    author = "Sarti, V. Mantovani and Feijoo, A. and Vida{\~n}a, I. and Ramos, A. and Giacosa, F. and Hyodo, T. and Kamiya, Y.",
    title = "{Constraining the low-energy S=-2 meson-baryon interaction with two-particle correlations}",
    eprint = "2309.08756",
    archivePrefix = "arXiv",
    primaryClass = "hep-ph",
    doi = "10.1103/PhysRevD.110.L011505",
    journal = "Phys. Rev. D",
    volume = "110",
    number = "1",
    pages = "L011505",
    year = "2024"
}

@article{Albaladejo:2024lam,
    author = "Albaladejo, M. and Feijoo, A. and Nieves, J. and Oset, E. and Vida{\~n}a, I.",
    title = "{Femtoscopy correlation functions and mass distributions from production experiments}",
    eprint = "2410.08880",
    archivePrefix = "arXiv",
    primaryClass = "hep-ph",
    doi = "10.1103/PhysRevD.110.114052",
    journal = "Phys. Rev. D",
    volume = "110",
    number = "11",
    pages = "114052",
    year = "2024"
}

@article{Torres-Rincon:2023qll,
    author = "Torres-Rincon, Juan M. and Ramos, {\`A}ngels and Tolos, Laura",
    title = "{Femtoscopy of D mesons and light mesons upon unitarized effective field theories}",
    eprint = "2307.02102",
    archivePrefix = "arXiv",
    primaryClass = "hep-ph",
    doi = "10.1103/PhysRevD.108.096008",
    journal = "Phys. Rev. D",
    volume = "108",
    number = "9",
    pages = "096008",
    year = "2023"
}

@article{Ikeno:2023ojl,
    author = "Ikeno, Natsumi and Toledo, Genaro and Oset, Eulogio",
    title = "{Model independent analysis of femtoscopic correlation functions: An application to the Ds0{\textasteriskcentered}(2317)}",
    eprint = "2305.16431",
    archivePrefix = "arXiv",
    primaryClass = "hep-ph",
    doi = "10.1016/j.physletb.2023.138281",
    journal = "Phys. Lett. B",
    volume = "847",
    pages = "138281",
    year = "2023"
}

@article{Vidana:2023olz,
    author = "Vidana, I. and Feijoo, A. and Albaladejo, M. and Nieves, J. and Oset, E.",
    title = "{Femtoscopic correlation function for the Tcc(3875)+ state}",
    eprint = "2303.06079",
    archivePrefix = "arXiv",
    primaryClass = "hep-ph",
    doi = "10.1016/j.physletb.2023.138201",
    journal = "Phys. Lett. B",
    volume = "846",
    pages = "138201",
    year = "2023"
}

@article{Ge:2025put,
    author = "Ge, Duo-Lun and Liu, Zhi-Wei and Lu, Jun-Xu and Geng, Li-Sheng",
    title = "{Deuteron-deuteron interaction and correlation function}",
    eprint = "2502.18872",
    archivePrefix = "arXiv",
    primaryClass = "nucl-th",
    doi = "10.1103/ttrc-qhv5",
    journal = "Phys. Rev. C",
    volume = "112",
    number = "3",
    pages = "034003",
    year = "2025"
}

@article{Molina:2023oeu,
    author = "Molina, R. and Liu, Zhi-Wei and Geng, Li-Sheng and Oset, E.",
    title = "{Correlation function for the $a_0(980)$}",
    eprint = "2312.11993",
    archivePrefix = "arXiv",
    primaryClass = "hep-ph",
    doi = "10.1140/epjc/s10052-024-12694-w",
    journal = "Eur. Phys. J. C",
    volume = "84",
    number = "3",
    pages = "328",
    year = "2024"
}

@article{Liu:2022nec,
    author = "Liu, Zhi-Wei and Li, Kai-Wen and Geng, Li-Sheng",
    title = "{Strangeness S = {\ensuremath{-}}2 baryon-baryon interactions and femtoscopic correlation functions in covariant chiral effective field theory*}",
    eprint = "2201.04997",
    archivePrefix = "arXiv",
    primaryClass = "hep-ph",
    doi = "10.1088/1674-1137/ac988a",
    journal = "Chin. Phys. C",
    volume = "47",
    number = "2",
    pages = "024108",
    year = "2023"
}

@article{Liu:2023wfo,
    author = "Liu, Zhi-Wei and Lu, Jun-Xu and Liu, Ming-Zhu and Geng, Li-Sheng",
    title = "{Distinguishing the spins of Pc(4440) and Pc(4457) with femtoscopic correlation functions}",
    eprint = "2305.19048",
    archivePrefix = "arXiv",
    primaryClass = "hep-ph",
    doi = "10.1103/PhysRevD.108.L031503",
    journal = "Phys. Rev. D",
    volume = "108",
    number = "3",
    pages = "L031503",
    year = "2023"
}

@article{Liu:2023uly,
    author = "Liu, Zhi-Wei and Lu, Jun-Xu and Geng, Li-Sheng",
    title = "{Study of the DK interaction with femtoscopic correlation functions}",
    eprint = "2302.01046",
    archivePrefix = "arXiv",
    primaryClass = "hep-ph",
    doi = "10.1103/PhysRevD.107.074019",
    journal = "Phys. Rev. D",
    volume = "107",
    number = "7",
    pages = "074019",
    year = "2023"
}

@article{Kamiya:2019uiw,
    author = "Kamiya, Yuki and Hyodo, Tetsuo and Morita, Kenji and Ohnishi, Akira and Weise, Wolfram",
    title = "{$K^-p$ Correlation Function from High-Energy Nuclear Collisions and Chiral SU(3) Dynamics}",
    eprint = "1911.01041",
    archivePrefix = "arXiv",
    primaryClass = "nucl-th",
    doi = "10.1103/PhysRevLett.124.132501",
    journal = "Phys. Rev. Lett.",
    volume = "124",
    number = "13",
    pages = "132501",
    year = "2020"
}

@article{Garcia-Recio:2002yxy,
    author = "Garcia-Recio, C. and Nieves, J. and Ruiz Arriola, E. and Vicente Vacas, M. J.",
    title = "{S = -1 meson baryon unitarized coupled channel chiral perturbation theory and the S(01) Lambda(1405) and Lambda(1670) resonances}",
    eprint = "hep-ph/0210311",
    archivePrefix = "arXiv",
    doi = "10.1103/PhysRevD.67.076009",
    journal = "Phys. Rev. D",
    volume = "67",
    pages = "076009",
    year = "2003"
}

@article{Oller:2006jw,
    author = "Oller, Jose A.",
    title = "{On the strangeness -1 S-wave meson-baryon scattering}",
    eprint = "hep-ph/0603134",
    archivePrefix = "arXiv",
    doi = "10.1140/epja/i2006-10011-3",
    journal = "Eur. Phys. J. A",
    volume = "28",
    pages = "63--82",
    year = "2006"
}

@article{Bruns:2010sv,
    author = "Bruns, Peter C. and Mai, Maxim and Mei{\ss}ner, Ulf-G.",
    title = "{Chiral dynamics of the S11(1535) and S11(1650) resonances revisited}",
    eprint = "1012.2233",
    archivePrefix = "arXiv",
    primaryClass = "nucl-th",
    reportNumber = "HISKP-TH-10-28, FZJ-IKP-TH-2010-25",
    doi = "10.1016/j.physletb.2011.02.008",
    journal = "Phys. Lett. B",
    volume = "697",
    pages = "254--259",
    year = "2011"
}

\clearpage
\begin{widetext}
\appendix
\label{appendix}

\section{Explicit expressions of the off-shell meson-baryon scattering amplitudes}
\label{app:amplitudes}

In this appendix, we collect the $S$-wave projected off-shell amplitudes used in the coupled-channel calculation. Throughout, the subscripts $i$($j$) and $a$($b$) label the initial (final) meson and baryon species, respectively, and $c$ denotes the intermediate exchanged octet baryon. The SU(3) coupling coefficients $C_{jb,c}^{\mathrm{Born}}$, $C_{jb,ia}^{\mathrm{NLO1}}$, and $C_{jb,ia}^{\mathrm{NLO2}}$ can be found in Refs.~\cite{Borasoy:2004kk,Borasoy:2005ie,Borasoy:2005fq,Oller:2005ig,Oller:2006jw}.

From the Yukawa vertex in Eq.~(\ref{eq:LOP1}), we derive the $s$-channel Born amplitude
\begin{equation}
    -i\mathcal{V}^{\mathrm{s\text{-}Born}}_{jb,ia}=\frac{1}{12f_i f_j}\sum_{c=1}^8C_{jb,c}^{\mathrm{Born}}C_{ia,c}^{\mathrm{Born}}\Bar{u}\left(p_b\right)\left(-\slashed{p}_j\gamma_{5}\right)\frac{i\left(\slashed{P}+M_c\right)}{s-M_c^2}\left(\slashed{p}_i\gamma_{5}\right)u\left(p_a\right).
\end{equation}
Evaluating the spinor algebra in the c.o.m.\ frame and performing the $S$-wave projection, with the shorthand $\Delta_{\pm}^{x} \equiv \sqrt{s} \pm M_x + E_x - p_x^0$ ($x = a, b$), yields
\begin{equation}
    \mathcal{V}^{\mathrm{s\text{-}Born}}_{jb,ia;0}\left(\sqrt{s},\mathrm{p}_b,\mathrm{p}_a,p_b^0,p_a^0\right)=\frac{N_bN_a}{12f_i f_j}\sum_{c=1}^8C_{jb,c}^{\mathrm{Born}}C_{ia,c}^{\mathrm{Born}} 
    \frac{\Delta_{-}^{b}\,\Delta_{-}^{a}}{\sqrt{s}+M_c}.
\end{equation}

The $u$-channel crossed Born diagram reads
\begin{equation}
    -i\mathcal{V}^{\mathrm{u\text{-}Born}}_{jb,ia}=\frac{1}{12f_i f_j}\sum_{c=1}^8C_{ic,b}^{\mathrm{Born}}C_{jc,a}^{\mathrm{Born}}\Bar{u}\left(p_b\right)\left(\slashed{p}_i\gamma_{5}\right)\frac{i\left(\slashed{P}+M_c\right)}{u-M_c^2}\left(-\slashed{p}_j\gamma_{5}\right)u\left(p_a\right),
\end{equation}
in which $P=p_a-p_j=p_b-p_i$ and the Mandelstam variable $u$ in the c.o.m.\ frame for the general off-shell case is
\begin{equation}
    u=\left(p_a-p_j\right)^2=\left(p_a^0+p_b^0-\sqrt{s}\right)^2-\left(\mathbf{p}_a+\mathbf{p}_b\right)^2.
\end{equation}
To present the $S$-wave projected result compactly, we introduce the auxiliary functions
\begin{align}
    \mathcal{F}_{+}&= \left(M_a + M_c\right) \left(M_b + M_c\right) \left(M_a + M_b - M_c + \sqrt{s}\right) \nonumber\\
    &\quad  - E_{b}^{2} \left(M_{a} + M_{c} - p_{a}^{0}\right) - M_{b}^{2} p_{a}^{0} - M_{b} M_{c} p_{a}^{0} + M_{b} \left(p_{a}^{0}\right)^{2} + M_{c} \left(p_{a}^{0}\right)^{2} \nonumber\\
    &\quad  - E_{a}^{2} \left(E_{b} + M_{b} + M_{c} - p_{b}^{0}\right) - M_{a}^{2} p_{b}^{0} - M_{a} M_{c} p_{b}^{0} + M_{c} p_{a}^{0} p_{b}^{0} \nonumber\\
    &\quad  - \left(p_{a}^{0}\right)^{2} p_{b}^{0} + M_{a} \left(p_{b}^{0}\right)^{2} + M_{c} \left(p_{b}^{0}\right)^{2} - p_{a}^{0} \left(p_{b}^{0}\right)^{2} \nonumber\\
    &\quad  + E_{b} \left(M_{a}^{2} - \left(M_{c} - p_{a}^{0}\right) \left(p_{a}^{0} - \sqrt{s}\right) + M_{a} \left(M_{c} + \sqrt{s}\right)\right) \nonumber\\
    &\quad  + E_{a} \left(-E_{b}^{2} + M_{b}^{2} - \left(M_{c} - p_{b}^{0}\right) \left(p_{b}^{0} - \sqrt{s}\right) + E_{b} \left(M_{c} + \sqrt{s}\right) + M_{b} \left(M_{c} + \sqrt{s}\right)\right) \nonumber\\
    &\quad  - M_{b} p_{a}^{0} \sqrt{s} - M_{c} p_{a}^{0} \sqrt{s} - M_{a} p_{b}^{0} \sqrt{s} - M_{c} p_{b}^{0} \sqrt{s} + p_{a}^{0} p_{b}^{0} \sqrt{s},
    \label{eq:Fplus}
\end{align}
\begin{align}
    \mathcal{F}_{-}&= \left(M_a + M_c\right) \left(M_b + M_c\right) \left(\sqrt{s} + M_c - M_a - M_b\right) \nonumber\\
    &\quad - M_{b}^{2} p_{a}^{0} - M_{b} M_{c} p_{a}^{0} - M_{b} \left(p_{a}^{0}\right)^{2} - M_{c} \left(p_{a}^{0}\right)^{2} + E_{b}^{2} \left(M_{a} + M_{c} + p_{a}^{0}\right) \nonumber \\
    &\quad  - M_{a}^{2} p_{b}^{0} - M_{a} M_{c} p_{b}^{0} - M_{c} p_{a}^{0} p_{b}^{0} - \left(p_{a}^{0}\right)^{2} p_{b}^{0} - M_{a} \left(p_{b}^{0}\right)^{2} - M_{c} \left(p_{b}^{0}\right)^{2} \nonumber \\
    &\quad  - p_{a}^{0} \left(p_{b}^{0}\right)^{2} + E_{a}^{2} \left(-E_{b} + M_{b} + M_{c} + p_{b}^{0}\right) \nonumber \\
    &\quad  + E_{b} \left(M_{a}^{2} + M_{a} \left(M_{c} - \sqrt{s}\right) + \left(M_{c} + p_{a}^{0}\right) \left(p_{a}^{0} - \sqrt{s}\right)\right) \nonumber \\
    &\quad  + E_{a} \left(-E_{b}^{2} + M_{b}^{2} + M_{b} \left(M_{c} - \sqrt{s}\right) + \left(M_{c} + p_{b}^{0}\right) \left(p_{b}^{0} - \sqrt{s}\right) + E_{b} \left(-M_{c} + \sqrt{s}\right)\right) \nonumber \\
    &\quad  + M_{b} p_{a}^{0} \sqrt{s} + M_{c} p_{a}^{0} \sqrt{s} + M_{a} p_{b}^{0} \sqrt{s} + M_{c} p_{b}^{0} \sqrt{s} + p_{a}^{0} p_{b}^{0} \sqrt{s},
    \label{eq:Fminus}
\end{align}
and
\begin{equation}
    B = \left(p_a^0+p_b^0-\sqrt{s}\right)^2-\mathrm{p}_a^2-\mathrm{p}_b^2-M_c^2.
\end{equation}
The $S$-wave projected amplitude then reads
\begin{align}
&\mathcal{V}^{\mathrm{u\text{-}Born}}_{jb,ia;0}\left(\sqrt{s},\mathrm{p}_b,\mathrm{p}_a,p_b^0,p_a^0\right) \nonumber\\
=&\;\frac{N_bN_a}{12f_j f_i}\sum_{c=1}^8C_{ic,b}^{\mathrm{Born}}C_{jc,a}^{\mathrm{Born}}\Bigg\{ 
\left(M_c+\sqrt{s}\right)-\frac{\mathcal{F}_{+}}{2\left(E_b+M_b\right)\left(E_a+M_a\right)} \nonumber \\
& + \bigg[ \mathcal{F}_{-} + \frac{B\,\mathcal{F}_{+}}{2\left(E_b+M_b\right)\left(E_a+M_a\right)} \bigg]
\times \frac{1}{4\mathrm{p}_a\mathrm{p}_b}\ln{\frac{B+2\mathrm{p}_a\mathrm{p}_b}{B-2\mathrm{p}_a\mathrm{p}_b}}\Bigg\}.
\label{eq:uBorn_Swave}
\end{align}

The $S$-wave projected NLO contact amplitude from the $\mathcal{O}\left(p^2\right)$ chiral Lagrangian in Eq.~(\ref{eq:LOP2}) reads
\begin{equation}
    \mathcal{V}^{\mathrm{NLO}}_{jb,ia;0}\left(\sqrt{s},\mathrm{p}_b,\mathrm{p}_a,p_b^0,p_a^0\right)
    =
    \frac{N_bN_a}{f_if_j}\left(C_{jb,ia}^{\mathrm{NLO1}}-2\,C_{jb,ia}^{\mathrm{NLO2}}\left(p_i^0p_j^0+\frac{\mathrm{p}_b^2\mathrm{p}_a^2}{3N_b^2N_a^2}\right)\right)
    ,
\end{equation}
in which $p_i^0=\sqrt{s}-p_a^0$, $p_j^0=\sqrt{s}-p_b^0$.

\section{Extracted parameters in the on-shell and off-shell schemes}
\label{app:paras}

\begin{table*}[htbp]
\centering
\caption{Best-fit parameters for the $S=-1$ sector: comparison between off-shell and on-shell schemes at WT, alongside the full LO results for the on-shell case. The table lists the $\chi^2/\text{d.o.f.}$, the common cutoff $\Lambda$ [GeV] used for all coupled channels, and the decay constant $f_{\mathrm{decay}}$ [MeV].}
\label{tab:parameters_LO}
\begin{tabular*}{\textwidth}{@{\extracolsep{\fill}}lccc}
\hline\hline
Model & $\chi^2/\text{d.o.f.}$ & $\Lambda$ [GeV] & $f_{\mathrm{decay}}$ [MeV]\\
\hline
WT-OFF   & 4.11 & 0.720 & 111.1 \\
WT-ON    & 3.48 & 0.635 & 109.4  \\
LO-ON    & 3.18 & 0.669 & 108.8  \\
\hline\hline
\end{tabular*}
\end{table*}
\begin{table*}[htbp]
\centering
\caption{Best-fit parameters for the $S=-1$ sector: comparison between off-shell and on-shell schemes at NLO. The top section lists the $\chi^2/\text{d.o.f.}$, the six cutoffs $\Lambda_j$ [GeV], and the decay constant $f_{\mathrm{decay}}$ [MeV]. The bottom section lists the NLO low-energy constants $b_i$ [GeV$^{-1}$].}
\label{tab:parameters_NLO}
\begin{tabular*}{\textwidth}{@{\extracolsep{\fill}}lcccccccc}
\hline\hline
Model & $\chi^2/\text{d.o.f.}$ & $\Lambda_{\bar{K}N}$ & $\Lambda_{\pi\Lambda}$ & $\Lambda_{\pi\Sigma}$ & $\Lambda_{\eta\Lambda}$ & $\Lambda_{\eta\Sigma}$ & $\Lambda_{K\Xi}$ & $f_{\mathrm{decay}}$ \\
\hline
NLO-OFF     & 1.98 & 0.652 & 1.401 & 0.866 & 0.859 & 0.304 & 0.481 & 121.7 \\
NLO-ON      & 1.82 & 0.524 & 0.323 & 0.654 & 0.755 & 0.669 & 0.228 & 113.8 \\
\hline
Model  & $b_0$ & $b_D$ & $b_F$ & $b_1$ & $b_2$ & $b_3$ & $b_4$ & \\
\hline
NLO-OFF    & -0.160 & -0.087 & 0.077 & -0.004 & -0.064 & 0.104 & 0.194 & \\
NLO-ON     & 0.186 & -0.506 & 0.207 & -0.173 & -0.201 & 0.153 & 0.264 & \\
\hline\hline
\end{tabular*}
\end{table*}

In Table~\ref{tab:parameters_LO} and Table~\ref{tab:parameters_NLO}, we present the best-fit parameter sets obtained from the on-shell and off-shell schemes at WT~(LO) and NLO, respectively. At WT, the results for both off-shell and on-shell cases are remarkably similar, a feature further reflected in the overlapping profiles of the corresponding CFs shown in Fig.~\ref{fig:CFR1p0}(a). This consistency is achieved through the fine-tuning of the cutoff $\Lambda$ and the decay constant $f_{\mathrm{decay}}$, which effectively renormalize the off-shell effects into these parameters. 

A comparison between the LO-ON and WT-ON results shows a marginal improvement in the $\chi^2/\text{d.o.f.}$, confirming that the Born terms in the on-shell formulation act as a minor correction to the leading WT interaction~\cite{Oset:1997it}. However, this trend does not persist in the off-shell framework. In fact, the full LO off-shell fit yields a significantly poorer description of the data compared to the WT-only case.
Note that high-energy scattering data from $\sigma(K^-p\rightarrow \eta\Lambda)$ are excluded in the fit at WT and LO, while included only at NLO~\cite{Ikeda:2012au,Guo:2012vv}, as shown in Fig.~\ref{fig:eta_lambda}.

\begin{figure}[htbp]
    \centering
    \includegraphics[width=0.48\textwidth]{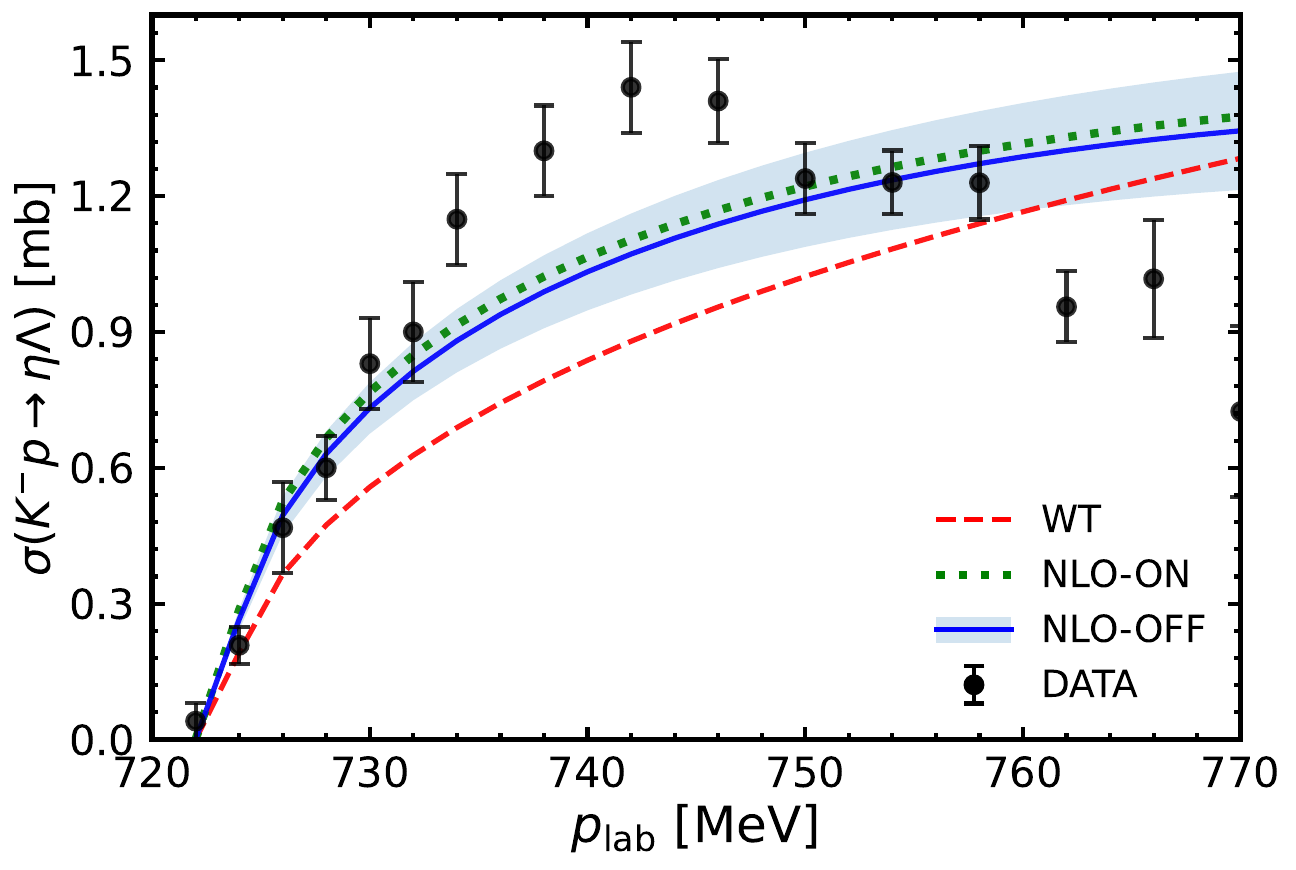}
    \caption{Total cross section for $K^-p \to \eta\Lambda$ as a function of the laboratory momentum $p_{\mathrm{lab}}$. The red dashed line represents the WT result, the blue solid line shows the NLO off-shell result, while the green dotted line corresponds to the NLO on-shell result for comparison. The shaded band indicates the NLO truncation uncertainty. Experimental data (black points) are taken from Ref.~\cite{CrystalBall:2001uhc}.}
    \label{fig:eta_lambda}
\end{figure}

Turning now to the NLO results, both the off-shell and on-shell formulations provide a comparably high-quality fit to the experimental data. For both schemes, the extracted cutoff parameters generally fall within the physically expected range, and the LECs remain at a consistent order of magnitude.

It is important to discuss the numerical performance of different regularization schemes. Although the on-shell approach within dimensional regularization~(DR) may yield a smaller $\chi^2/\mathrm{d.o.f.}$ compared to the cutoff regularization (in both on-shell and off-shell schemes), this numerical superiority often stems from excessive fine-tuning of the subtraction constants $a(\mu)$~\cite{Ikeda:2012au,Guo:2012vv,Mai:2012dt}. 

In a physically natural scenario, the subtraction constant is expected to be around $a(\mu) \approx -2$ (at a renormalization scale $\mu = 630$~MeV), and its absolute value should not significantly exceed unity to remain consistent with the typical scales of chiral symmetry breaking~\cite{Oller:2000fj}. This requirement for parameter ``naturalness" is a primary motivation for our preference for the cutoff regularization scheme~\cite{Epelbaum:2014efa}. By employing a cutoff scale $\Lambda$ that corresponds to the realistic physical cutoff interval of the effective field theory, we ensure that the model parameters remain within a well-defined, physically meaningful domain, thereby avoiding potential artifacts arising from unconstrained numerical minimization.

\section{Global fit results in the on-shell scheme}
\label{app:fit_on}

\begin{table*}[htbp]
\centering
\caption{On-shell results for the threshold ratios, scattering lengths, $\chi^2/\text{d.o.f.}$, and the two $I=0$ pole positions, shown in a format identical to the off-shell results in Table~\ref{tab:results_off} except for the extra LO case.}
\label{tab:results_on}
\renewcommand{\arraystretch}{1.4}

\begin{tabular*}{\linewidth}{@{\extracolsep{\fill}} l c c c c c }
\hline\hline
& $\chi^2/\text{d.o.f}$ & $a_{K^-p}$ [fm] & $\gamma$ & $R_c$ & $R_n$ \\
\hline
NLO & 1.82 & $(-0.81 \pm 0.08) + i(1.00 \pm 0.10)$ & $2.36 \pm 0.24$ & $0.662 \pm 0.063$ & $0.193 \pm 0.025$ \\
LO & 3.18 & $-0.55 + i(0.93)$ & $2.09$ & $0.617$ & $0.320$ \\
WT & 3.48 & $-0.81 + i(1.00)$ & $2.35$ & $0.622$ & $0.244$ \\
EXP & -- & $(-0.65 \pm 0.10) + i(0.81 \pm 0.15)$ & $2.36 \pm 0.04$ & $0.664 \pm 0.011$ & $0.189 \pm 0.015$ \\
\hline
\end{tabular*}

\begin{tabular*}{\linewidth}{@{\extracolsep{\fill}} l c c c c c }
& Pole positions [MeV] & $|g_{\bar{K}N}|$ [GeV] & $|g_{\pi\Sigma}|$ [GeV] & $|g_{\eta\Lambda}|$ [GeV] & $|g_{K\Xi}|$ [GeV] \\
\hline
NLO $\Lambda(1380)$ & $1376 \pm 3 - i(61 \pm 1)$ & $8.56 \pm 0.85$ & $7.69 \pm 0.12$ & $1.83 \pm 0.13$ & $1.72 \pm 0.04$ \\
NLO $\Lambda(1405)$ & $1432 \pm 1 - i(27 \pm 2)$ & $7.87 \pm 0.32$ & $5.22 \pm 0.21$ & $4.32 \pm 0.16$ & $1.33 \pm 0.07$ \\
LO $\Lambda(1380)$ & $1387 - i(43)$ & $7.49$ & $7.23$ & $3.15$ & $0.68$ \\
LO $\Lambda(1405)$~(virtual) & $1444 - i(40)$ & $8.20$ & $6.98$ & $5.14$ & $1.09$ \\
WT $\Lambda(1380)$ & $1385 - i(58)$ & $5.89$ & $7.32$ & $2.25$ & $1.61$ \\
WT $\Lambda(1405)$ & $1429 - i(20)$ & $6.88$ & $4.57$ & $3.82$ & $1.10$ \\
\hline\hline
\end{tabular*}

\end{table*}

\begin{figure*}[htpb]
    \centering
    \includegraphics[width=7.0in]{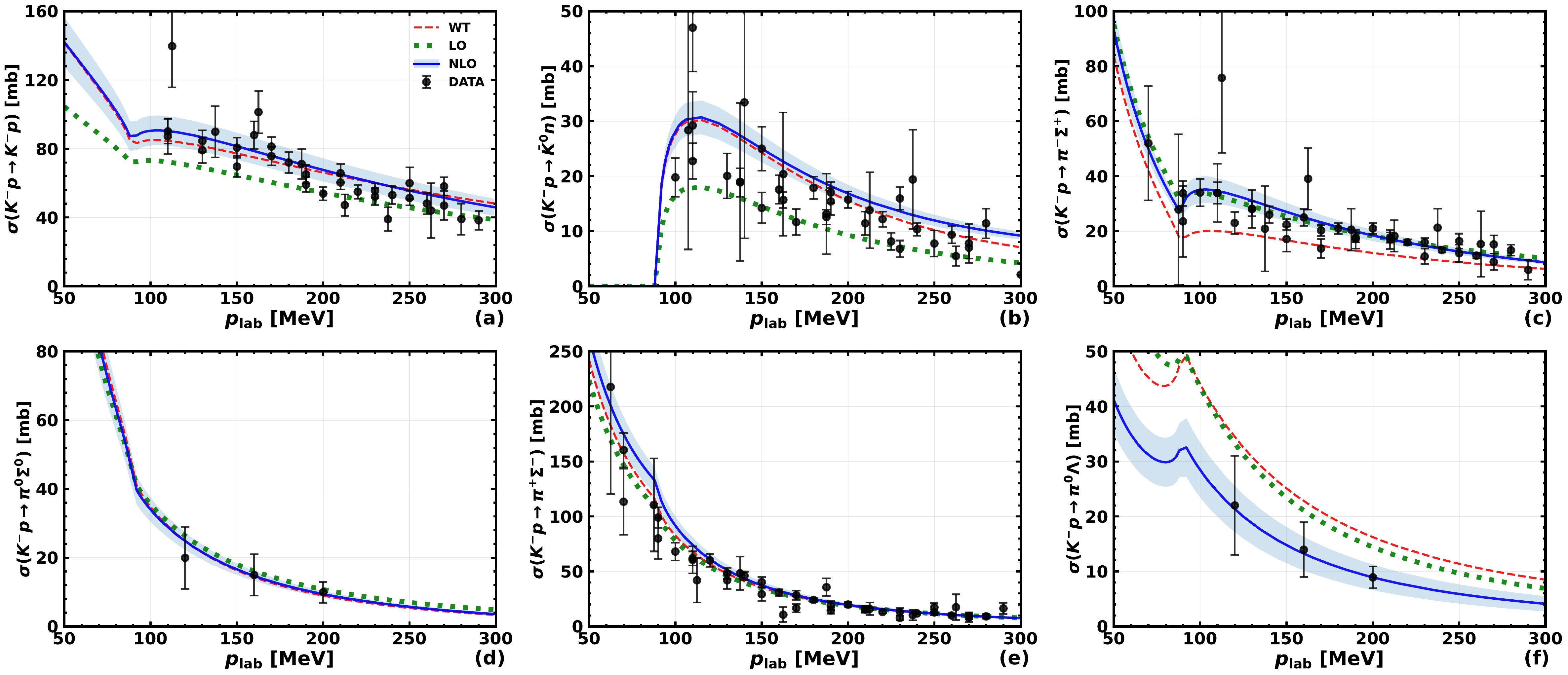}
    \caption{Global fit results for $K^-p$ scattering and subthreshold dynamics in the on-shell scheme.
    Same as Fig.~\ref{fig:obs_off} but for the on-shell case. The results at WT (red dashed), LO (green dotted), and NLO (blue solid) levels are compared with experimental data.}
    \label{fig:obs_on}
\end{figure*}

In Table~\ref{tab:results_on}, we present the on-shell results for the threshold ratios, scattering lengths, $\chi^2/\text{d.o.f.}$, and the positions of the two $I=0$ poles, with the corresponding global fit results shown in Fig.~\ref{fig:obs_on}. The layout follows the format of the off-shell results in Table~\ref{tab:results_off}, with the addition of the full LO case for completeness. A notable feature is the clear convergence pattern observed when moving from WT to NLO, particularly for the observables related to the $K^-p$ channel. 

 While the inclusion of the Born terms at LO yields a modest improvement in the overall $\chi^2/\text{d.o.f.}$ compared to the WT-only fit, the resulting pole structure appears less physically favorable. Specifically, the higher-mass pole in the LO case emerges as a virtual state located above the $\bar{K}N$ threshold, a configuration that deviates from the expected two-pole behavior observed in the more comprehensive NLO analysis.

\end{widetext}

\end{document}